\documentclass[letterpaper,final,onecolumn,accepted=2026-07-02]{quantumarticle}

\pdfoutput=1

\usepackage[utf8]{inputenc}
\usepackage[english]{babel}
\usepackage[dvipsnames,table,xcdraw]{xcolor}

\usepackage{newfloat}
\DeclareFloatingEnvironment[fileext=loa,
  listname={List of Algorithms},
  name=Algorithm, placement=htbp]{algorithm}

\usepackage[noend]{algpseudocode}
\usepackage{amsmath, amssymb, amsthm, mathdots, mathrsfs, mathtools}
\usepackage{array, bbm, booktabs, enumitem, tablefootnote, bbold, graphicx, hyperref}
\usepackage[T1]{fontenc}
\usepackage{natbib}
\usepackage{pgfplots, tikz}
\usetikzlibrary{arrows}
\usetikzlibrary{patterns}
\pgfplotsset{compat=1.15}
\usepackage{subcaption, multicol, multirow}

\newcommand{\customnameref}[2][black]{%
  \hypersetup{linkcolor=#1} 
  \nameref{#2} 
  \hypersetup{linkcolor=blue} 
}

\newtheorem{theorem}{Theorem}
\newtheorem{lemma}{Lemma}

\theoremstyle{definition}
\newtheorem{definition}{Definition}

\allowdisplaybreaks

\newcommand{\R}{\mathbb{R}}

\newcommand{\ov}{\overline}

\newcommand{\nHising}{N(H_{\textsc{Ising}})}
\newcommand{\nTotal}{N(\textsc{Total}_{\textsc{Ops}})}

\newcommand{\maxcut}{\delta^{\max}}
\newcommand{\hising}{$H_{\mathrm{Ising}}$ }

\usepackage[normalem]{ulem}
\usepackage{bm}
\usepackage{endnotes}

\usepackage{mathtools}
\DeclarePairedDelimiter\bra{\langle}{\rvert}
\DeclarePairedDelimiter\ket{\lvert}{\rangle}
\DeclarePairedDelimiterX\braket[2]{\langle}{\rangle}{#1\,\delimsize\vert\,\mathopen{}#2}

\newcommand{\unionofstars}{\textsc{union-of-stars}}

\DeclareUnicodeCharacter{2009}{\,} 

\author{Jai Moondra}
\affiliation{Carnegie Mellon University}
\email{jmoondra@andrew.cmu.edu}
\thanks{corresponding author. Work done while the author was at Georgia Institute of Technology.}

\author{Phillip C. Lotshaw}
\affiliation{Oak Ridge National Laboratory}

\author{Greg Mohler}
\affiliation{Georgia Tech Research Institute}

\author{Swati Gupta}
\affiliation{Massachusetts Institute of Technology}

\begin{document}

\title{Promise of Graph Sparsification and Decomposition for Noise Reduction in QAOA: Analysis for Trapped-Ion Compilations}

\maketitle

\begin{abstract}
We develop new approximate compilation schemes that significantly reduce the expense of compiling the Quantum Approximate Optimization Algorithm (QAOA) for solving the Max-Cut problem.  Our main focus is on compilation with trapped-ion simulators using Pauli-$X$ operations and all-to-all Ising Hamiltonian $H_\text{Ising}$ evolution generated by M{\o}lmer-S{\o}rensen or optical dipole force interactions, though some of our results also apply to standard gate-based compilations. Our results are based on principles of graph sparsification and decomposition; the former reduces the number of edges in a graph while maintaining its cut structure, while the latter breaks a weighted graph into a small number of unweighted graphs. Though these techniques have been used as heuristics in various hybrid quantum algorithms, there have been no guarantees on their performance, to the best of our knowledge. This work provides the first provable guarantees using sparsification and decomposition to improve quantum noise resilience and reduce quantum circuit complexity. 

For quantum hardware that uses edge-by-edge QAOA compilations, sparsification leads to a direct reduction in circuit complexity. For trapped-ion quantum simulators implementing all-to-all \hising pulses, we show that for a $(1-\epsilon)$ factor loss in the Max-Cut approximation ($\epsilon>0)$,  our compilations improve the (worst-case) number of \hising pulses from $O(n^2)$ to $O(n\log(n/\epsilon))$ and the (worst-case) number of Pauli-$X$ bit flips from $O(n^2)$ to $O\left(\frac{n\log(n/\epsilon)}{\epsilon^2}\right)$ for $n$-node graphs. This is an asymptotic improvement for any constant $\epsilon > 0$.
We demonstrate that significant improvements to the approximation ratio are obtained using decomposition in simulated trapped-ion experiments with dephasing noise. We further present a generic argument showing that sparsification results in an exponentially improved circuit fidelity lower bound in digital computing schemes based on one- and two-qubit gates, which are relevant to a wide variety of hardwares such as superconducting qubits and certain neutral atom or trapped ion setups, and more sophisticated noise models. We anticipate these approximate compilation techniques will be useful tools in a variety of future quantum computing experiments. 

\footnote{{This manuscript has been authored by UT-Battelle, LLC under Contract No. DE-AC05-00OR22725 with the U.S. Department of Energy. The United States Government retains and the publisher, by accepting the article for publication, acknowledges that the United States Government retains a non-exclusive, paid-up, irrevocable, world-wide license to publish or reproduce the published form of this manuscript, or allow others to do so, for United States Government purposes. The Department of Energy will provide public access to these results of federally sponsored research in accordance with the DOE Public Access Plan (http://energy.gov/downloads/doe-public-access-plan).}}

\end{abstract}

\section{Introduction}

\begin{figure*}[t]
    \centering
    \begin{subfigure}{0.32\textwidth}
        \includegraphics[width=\textwidth]{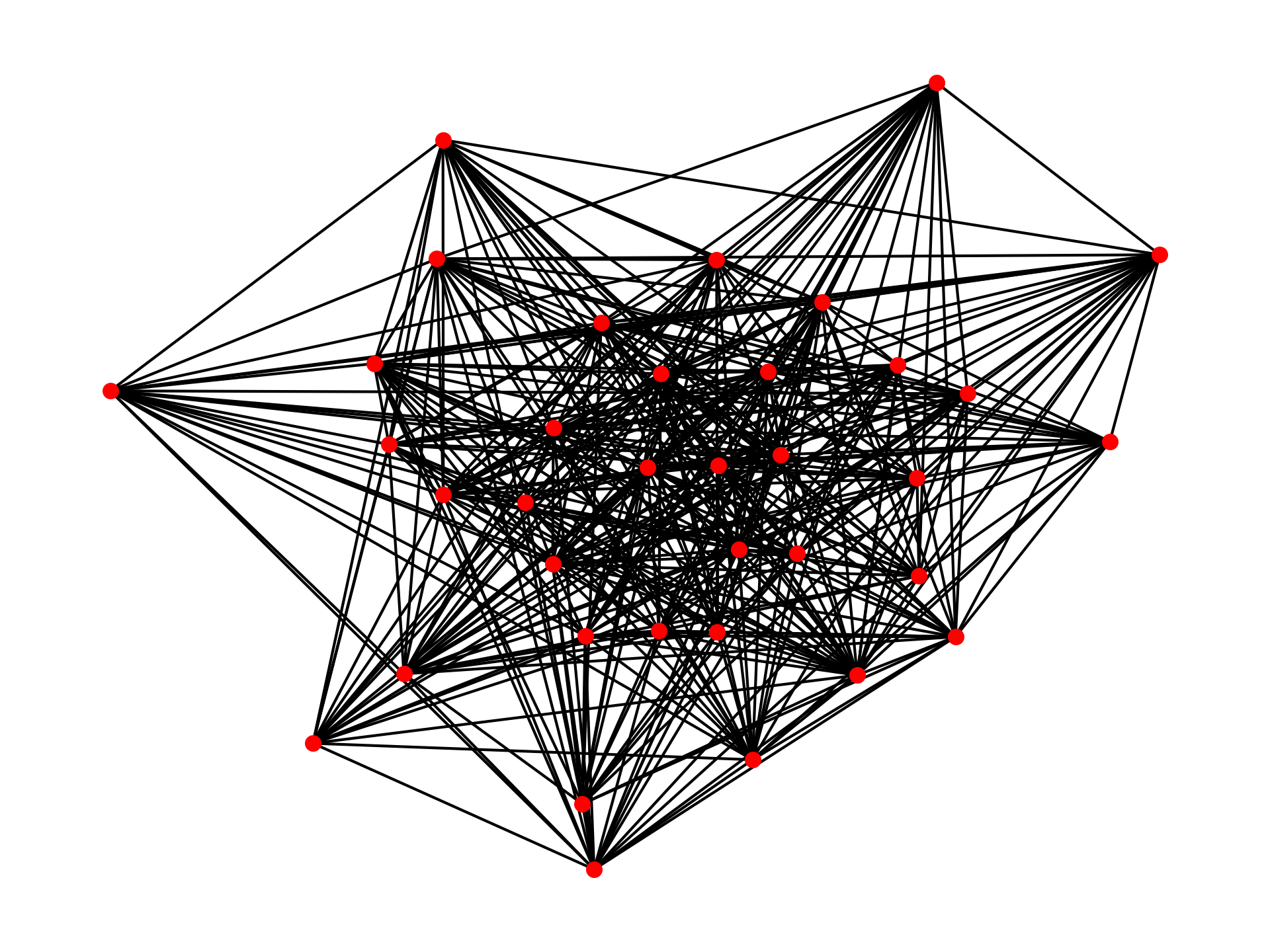}
    \end{subfigure}
    \hfill
    \begin{subfigure}{0.32\textwidth}
        \includegraphics[width=\textwidth]{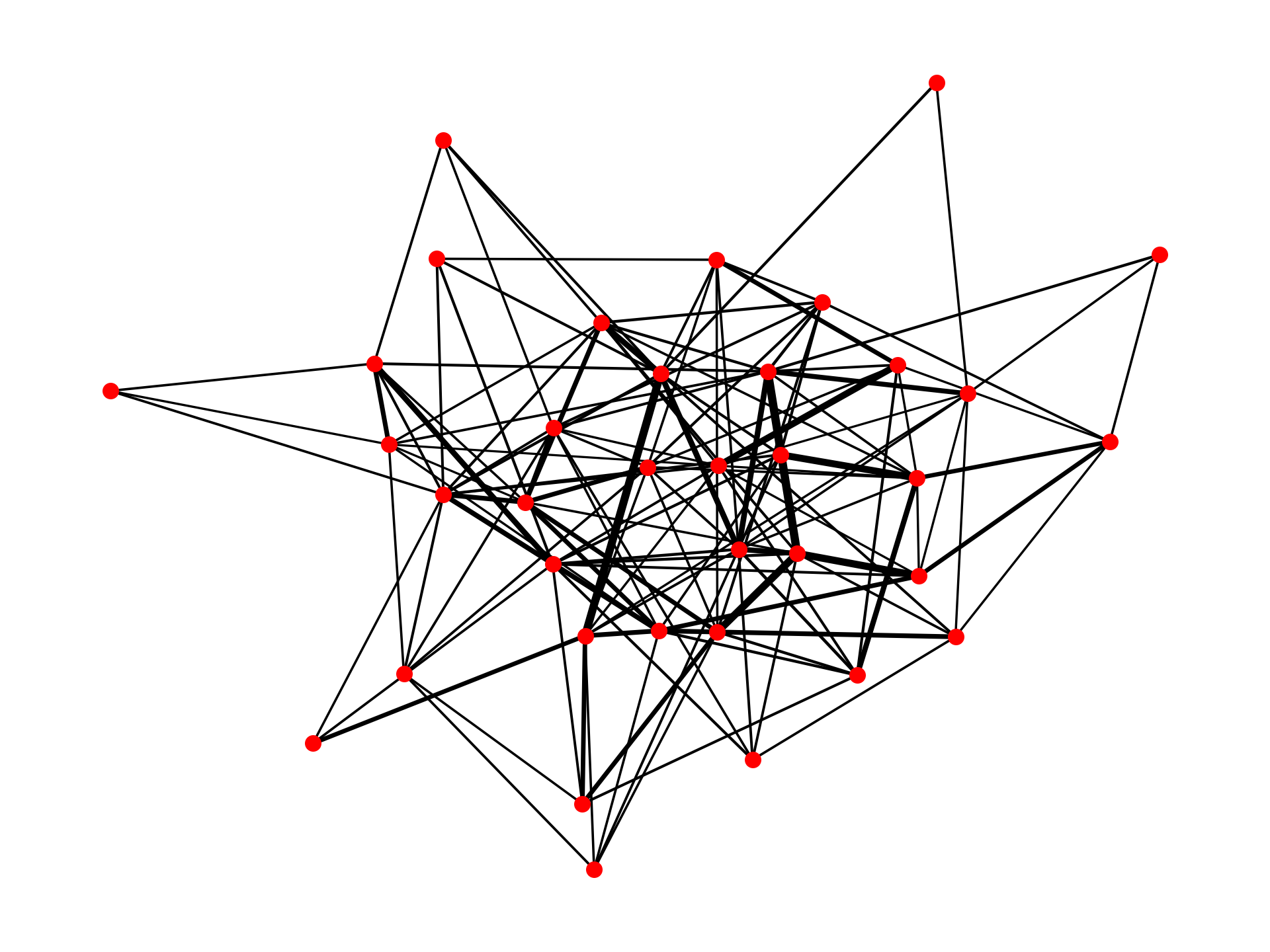}
    \end{subfigure}
    \hfill
    \begin{subfigure}{0.32\textwidth}
        \includegraphics[width=\textwidth]{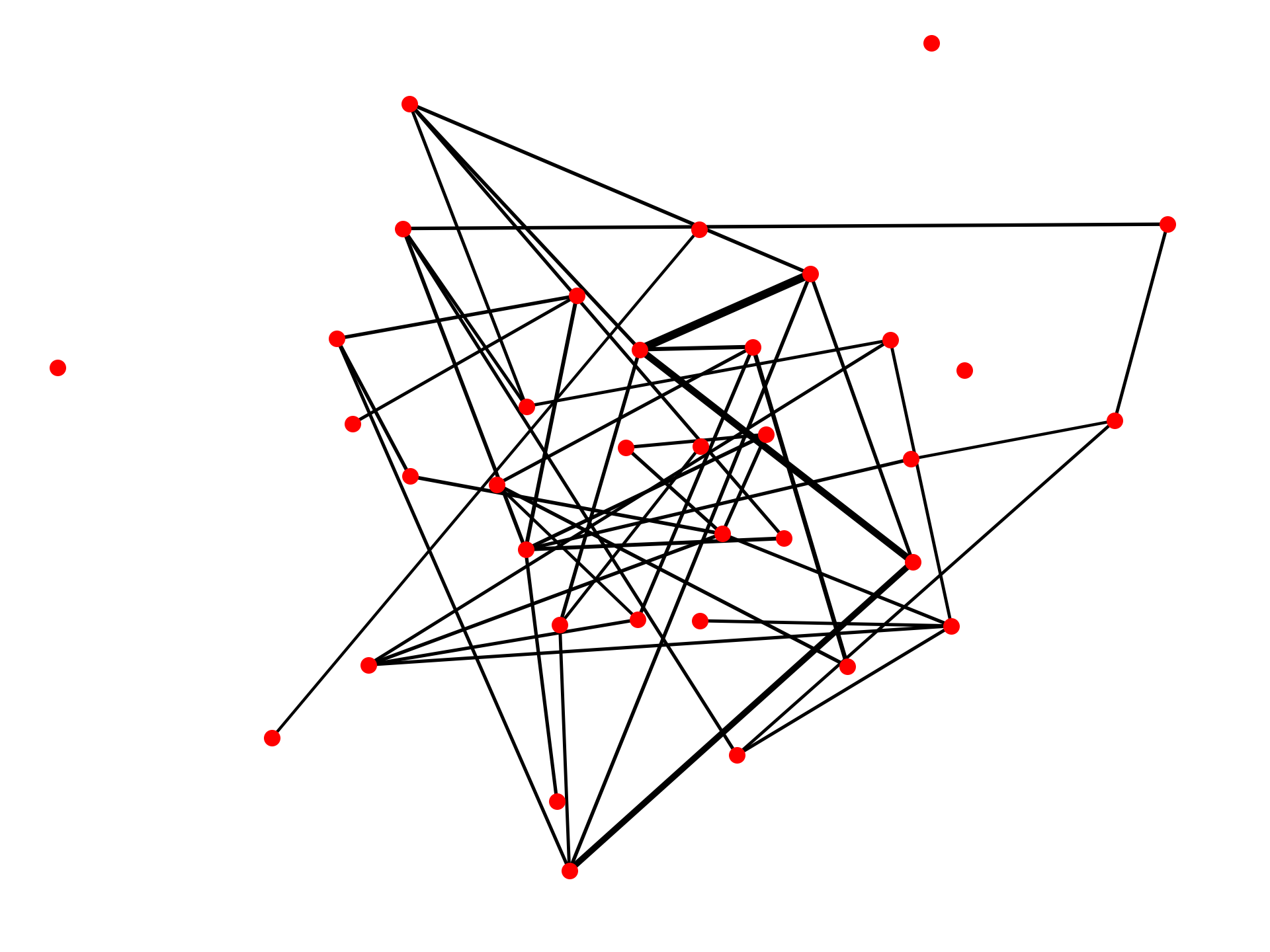}
    \end{subfigure}
    \caption{Left to right: graph sparsification, with the original unweighted graph $G$ (left, 397 edges) and two weighted graph sparsifiers $H_1$ (middle, 136 edges) and $H_2$ (right, 48 edges) of $G$ (all graphs have the same number of vertices). The Max-Cut in $H_1$ gives a cut in $G$ with cut value that is $90\%$ of the Max-Cut in $G$, i.e., is a $0.9$-approximation to Max-Cut in $G$. The Max-Cut in $H_2$ is a $0.82$-approximation to Max-Cut in $G$.
    }
    \label{fig: sparsification-example}
\end{figure*}

\begin{figure*}[t]
    \centering
    \includegraphics[width=0.8\textwidth]{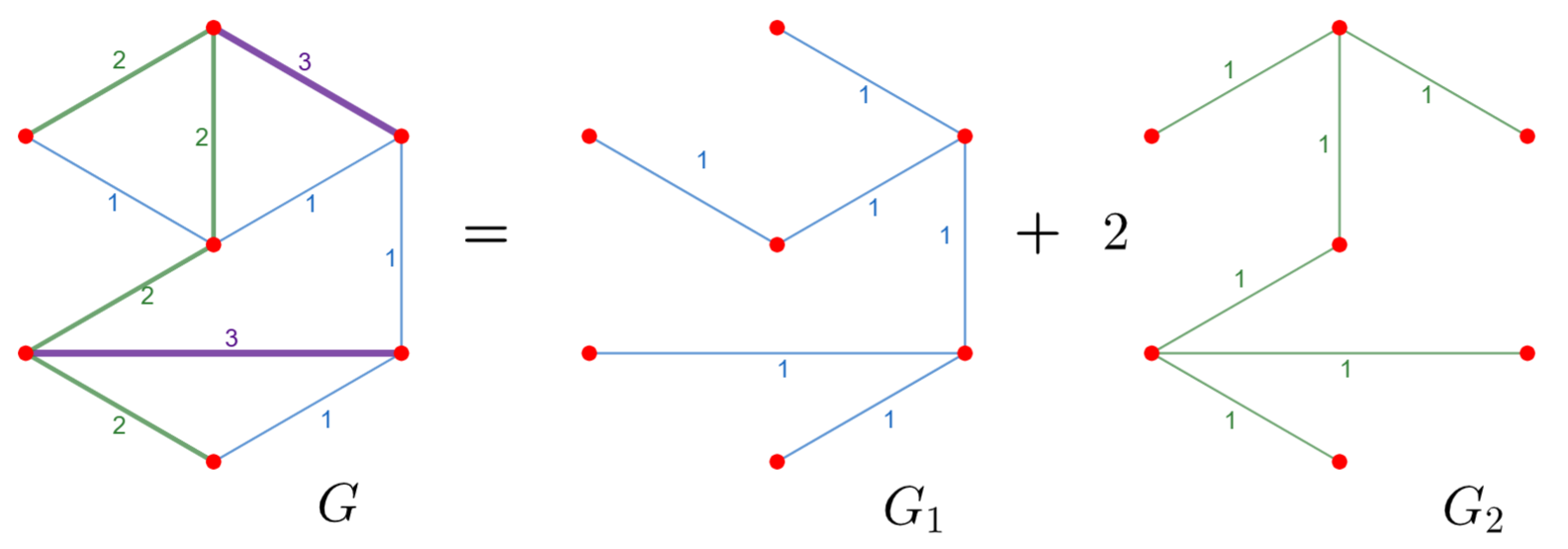}
    \caption{The graph decomposition process: The weighted graph $G$ on the left can be written as the weighted sum $G_1 +  2 G_2$ of two unweighted graphs $G_1, G_2$ on the right.}
    \label{fig: decomposition-example} 
\end{figure*}

The advantage of quantum computers over their classical counterparts for some computational tasks has been established for decades \cite{shor_scheme_1995, grover_fast_1996}. Promising new quantum algorithms for complex problems such as constrained optimization \cite{kerenidis_quantum_2020, augustino_quantum_2023, nannicini_fast_2024} and combinatorial optimization \cite{farhi_quantum_2014, peruzzo_variational_2014, tasseff_emerging_2022, sutter_quantum_2021, shaydulin_multistart_2019, harwood_formulating_2021, farhi_quantum_2020, farhi_quantum_2022,morris2024performant} are an active area of research. A leading benchmark for quantum optimization is the Max-Cut problem: given a graph $G = (V, E, c)$ on vertices $V$, edges $E$, and edge weights $c: E \to \R$, one seeks to find the subset $S \subseteq V$ maximizing the weighted cut value $\delta(S) 
:= \sum_{uv \in E: u \in S, v \in V \setminus S} c_{uv}$.
The development of the Quantum Approximate Optimization Algorithm (QAOA) \cite{farhi_quantum_2014} for Max-Cut has made it a leading candidate for demonstrating {quantum advantage} over classical computing. 

While classical algorithms currently have much faster runtimes, they are widely believed to have fundamental limitations when it comes to the solution quality, which is measured by the approximation ratio, defined as the ratio of the algorithm's cut value to the optimal cut value \cite{vazirani_approximation_2003, williamson_design_2010}. The state-of-the-art for Max-Cut is the Goemans-Williamson (GW) \cite{goemans_879-approximation_1994} algorithm with (tight \cite{karloff_how_1999}) approximation ratio $\alpha^* \simeq 0.878$ for all graphs with nonnegative edge weights. On the other hand, algorithms like QAOA are known to converge to the Max-Cut under the adiabatic limit (i.e., as the depth goes to infinity) \cite{farhi_quantum_2014}. Though known approximation bounds at finite circuit depths of QAOA do not outperform Goemans-Williamson's 0.878, computational advantages of QAOA relative to classical algorithms have been presented in varying contexts \cite{tate_warm-started_2023,basso2022quantum,farhi_quantum_2022,shaydulin2024evidence,zhou_quantum_2020}.

A significant bottleneck that hinders the attainment of quantum advantage with existing hardware is the loss in performance due to quantum noise arising from imperfect control or unwanted interactions with the environment. 
Limiting the impact of noise is therefore important both for progress \cite{ryan-anderson_realization_2021, clark_high-fidelity_2021} towards building fault-tolerant quantum computers~\cite{shor_scheme_1995, sivak_real-time_2023, krinner_realizing_2022} and for solving medium to large problem instances with near-term Noisy Intermediate-Scale Quantum (NISQ) computers~\cite{preskill_quantum_2018, dongmei_noisy_2025}. There has also been interest in using special-purpose quantum simulators with global addressing for quantum optimization at scales that are currently intractable with local one- and two-qubit gate addressing~\cite{rajakumar_generating_2022,ebadi2022quantum,pagano_quantum_2019}.

In this work, we approach quantum noise reduction from a \emph{problem-aware} perspective, and ask:

\begin{widetext}
\begin{center}
{\it Is it possible to classically modify the problem instance to improve quantum noise resilience while still being able to recover the correct solution? Are there provable guarantees for such techniques?}
\end{center} 
\end{widetext}

In the context of Max-Cut and QAOA, we answer these questions affirmatively using two classical pre-processing steps to reduce circuit complexity: graph sparsification and graph decomposition. Graph sparsification is a principled approach to reduce the number of edges $m = |E|$ in the graph and introduce weights on remaining edges, while ensuring that every cut in the original graph is approximated by the corresponding cut in the sparsified instance. A higher level of graph sparsification degrades the approximation guarantee; see the example in Fig.~\ref{fig: sparsification-example}. Sparsification has been used extensively in classical computing; however, its potential has not been well explored in quantum optimization. Recent work has used sparsifying heuristics to improve QAOA performance, but without theoretical guarantees~\cite{liu_quantum_2022}. 

Graph decomposition expresses a weighted graph as a weighted sum of a logarithmic (in the number of vertices, denoted $n = |V|$) number of unweighted graphs with controlled approximation error, as seen in Fig.~\ref{fig: decomposition-example}. Decomposition is tailored to a specific compilation technique \cite{rajakumar_generating_2022} with trapped-ion global gates, while sparsification is generally applicable to any QAOA implementation and may be especially beneficial for limiting SWAP gates in hardware with limited connectivity \cite{lotshaw2022scaling}. Both of these techniques provide more efficient compilations -- requiring less time and fewer steps for execution -- with less accumulated noise.

QAOA begins with a classical problem defined in terms of a cost function $\mathcal{C}(\bm z)$ with a bit string argument $\bm z = (z_1,\ldots z_N)$. This is mapped to a quantum \emph{cost Hamiltonian} $C$ with an eigenspectrum that contains the set of classical solutions $C\ket{\bm z} = \mathcal{C}(\bm z)\ket{\bm z}$, such that identifying the ground state of $C$ provides the optimal solution \cite{lucas2014ising}. To approximately solve these problems, QAOA evolves a quantum state in $p$ layers of Hamiltonian evolution, where each layer alternates between evolution under $C$ and under $B = \sum_i \sigma^x_i$ where $\sigma^x_i$ is the Pauli-$X$ operator on qubit $i$,
\begin{equation} \label{QAOA state} \ket{\bm \gamma,\bm \beta} = \prod_{l=1}^p e^{-i \beta_l B} e^{-i \gamma_l C} \ket{\psi_0}, \end{equation} 
where $\ket{\psi_0}$ is the initial state (i.e., $\ket{+}^{n}$ for standard QAOA, or a classically computed warm-start \citep{egger2021warm, tate_warm-started_2023}). The $\bm \gamma$ and $\bm \beta$ are variational parameters chosen to minimize $\langle C \rangle := \langle \bm z | C | \bm z \rangle$, such that measurements of the quantum state $\ket{\bm z}$ provide approximate solutions to the combinatorial problem.\footnote{Note we have chosen a convention in which we are beginning in the ground state of $-B$ and aiming to prepare the ground state of $+C$, or equivalently beginning in the highest state of $+B$ and trying to prepare the highest state of $-C$. The initial and target states are adiabatically connected in the limit $p\to \infty$ for the usual reasons \cite{farhi_quantum_2014}; see also Ref.~\cite{binkowski2024elementary} for a rigorous proof of convergence criteria under different choices of $B$ and $C$. We did not include the alternating signs directly in (\ref{QAOA state}) to better match the standard conventions, instead absorbing them into the parameters, which have arbitrary signs.} Given graph $G = (V, E, c)$ with edge costs $c_{e}$, the expected sum of costs of cut edges across a cut $\delta_G(S)$ (for $S\subseteq V$), is related to the graph coupling operator expectation as $\delta_G(S) = \frac{1}{2}\left(\sum_{e \in E} c_{e} - \langle {\bm z}_S | C | {\bm z}_S\rangle\right)$, where ${\bm z}_S$ is the corresponding bit string. Therefore, minimizing $\langle C \rangle$ is equivalent to finding $S \subseteq V$ with the maximum cut value $\delta_G(S)$. We discuss this further in Section \ref{sec: preliminaries}. In quantum computers with local-only addressing, each edge requires a noisy two-qubit gate operation \cite{lotshaw2022scaling} to compile $\exp(-i \gamma C)$. This case presents an obvious advantage for sparsified instances with fewer edges. For global-gate compilations more involved compilation schemes are required.

We focus on compilations for trapped-ion hardware with global  M{\o}lmer-S{\o}renson (MS) \cite{lotshaw_modeling_2023} or optical dipole force \cite{bohnet2016quantum} interactions. These gates natively implement an equally-weighted all-to-all Ising evolution with Hamiltonian 
\begin{equation}
    H_\text{Ising} = \sum_{i<j} \sigma_i^z\sigma_j^z.
\end{equation} \cite{rajakumar_generating_2022} show that arbitrary graph couplings for QAOA can be compiled on trapped-ion hardware using a sequence of global \hising pulses and individual ``bit flips'' (single-qubit Pauli-$X$ unitaries) to produce effective Ising interactions localized to star graphs, which can be summed to generate $C$ for any arbitrary graph. Their algorithm -- called \unionofstars\ -- compiles unweighted graphs with $O(n)$ \hising pulses while weighted graphs require $O(m) = O(n^2)$ \hising pulses in the worst-case; for weighted and unweighted graphs the number of bit flip operations is $O(m)$. Compilations that use fewer \hising pulses and bit flips can potentially reduce noise in the compiled circuit.

\begin{table*}[t]
\centering
\caption{Worst-case bounds on number of \hising pulses (denoted $\nHising$) and total number of operations (number of \hising pulses and bit flips, denoted $\nTotal$) on weighted and unweighted (dense) graphs on $|V| = n$ vertices and $|E| = m = O(n^2)$ edges. For any constant $\epsilon$, our algorithms yield an asymptotic improvement in the number of pulses and total number of operations, while keeping `large' or `non-trivial' cut values (see Section \ref{sec: preliminaries} for formal definition) within a factor $1 \pm \epsilon$. The best bounds (up to log factors) for weighted graphs are highlighted in yellow; note that using sparsified + decomposed graphs achieves the best bounds for both $\nHising$ and $\nTotal$ simultaneously.
}\label{tab: bounds-summary}
\begin{tabular}
{|c|c|c|c|c|c|c|}
\hline
Instance &
  Weighted? &
  \begin{tabular}[c]{@{}c@{}}Number\\ of edges\end{tabular} &
  \begin{tabular}[c]{@{}c@{}}Loss in cut\\ approximation 
  \end{tabular} &
  $\nHising$ &
  $\nTotal$ &
  Notes \\ \hline
\multirow{2}{*}{\begin{tabular}[c]{@{}c@{}}Original\\ Graph\end{tabular}} &
  $\times$ &
  $O(n^2)$ &
  Exact &
  $O(n)$ &
  $O(n^2)$ &
  \multirow{2}{*}{\cite{rajakumar_generating_2022}} \\ \cline{2-6}
 &
  $\checkmark$ &
  $O(n^2)$ &
  Exact &
  $O(n^2)$ &
  $O(n^2)$ &
   \\ \hline
\begin{tabular}[c]{@{}c@{}}Decomposed \\ Graph\end{tabular} &
  $\checkmark$ &
  $O(n^2)$ &
  $1 - \epsilon$ &
  \cellcolor{yellow!30} $O\left(n \log \frac{n}{\epsilon}\right)$ &
  $O(n^2)$ &
  Theorem \ref{thm: faster-union-of-stars-weighted-graphs} \\ \hline
\begin{tabular}[c]{@{}c@{}}Sparsified\\ Graph\end{tabular} &
  $\checkmark$ &
  $O\left(n/\epsilon^2\right)$ &
  $1 - \epsilon$ &
  $O\left(n/\epsilon^2\right)$ &
  \cellcolor{yellow!30} $O\left(n/\epsilon^2\right)$ & \begin{tabular}[c]{@{}c@{}}
  Follows from \\ \cite{batson_twice-ramanujan_2014} and \cite{rajakumar_generating_2022} \end{tabular} \\ \hline
\begin{tabular}[c]{@{}c@{}}Sparsified +\\ Decomposed\\ Graph\end{tabular} &
  $\checkmark$ &
  $O\left(n/\epsilon^2\right)$ &
  $1 - \epsilon$ &
  \cellcolor{yellow!30} $O\left(n \log\frac{n}{\epsilon}\right)$ &
  \cellcolor{yellow!30} $O\left(\frac{n \log(n/\epsilon)}{\epsilon^2}\right)$ &
  Theorem \ref{thm: second-graph-coupling-number} \\ \hline
\end{tabular}
\end{table*}

In our first contribution, we show that if a $1 - \epsilon$ factor approximation loss in Max-Cut approximation can be tolerated, then a weighted graph $G$ on $n$ vertices can be written as a (weighted) sum of at most $O(\log(n/\epsilon))$ unweighted graphs using our decomposition technique. Here, $\epsilon > 0$ is an arbitrary number that can be chosen to suit the needs of optimization. The smaller $\epsilon$ is, the better the Max-Cut approximation. Consequently, we can compile $G$ by composing the compilations of each of these unweighted graphs. This uses a total of $O(n \log(n/\epsilon))$ \hising pulses in the worst-case, which is asymptotically much smaller than the $O(m) = O(n^2)$ pulses needed by the edge-by-edge compilation in \unionofstars (e.g., $O(n \log n)$ vs $O(n^2)$ even for $\epsilon = 1/\log n$).

Next, we use classical graph sparsifiers to reduce the number of edges from $m$ to $O\left(\frac{n}{\epsilon^2}\right)$. Our sparsification approach is applicable to quantum hardware with local or global addressing; for global gate compilations, applying graph decomposition to the sparsified graph yields compilations that use at most $O\left(\frac{n\log(n/\epsilon)}{\epsilon^2}\right)$ bit flips, again asymptotically smaller than the $O(m) = O(n^2)$ bit flips used by \unionofstars. These bounds are summarized in Table \ref{tab: bounds-summary}. 
We verify them through simulations on graphs from the MQLib library \cite{dunning_what_2018}, which is a suite of graphs that serve as a benchmark for Max-Cut algorithms. {\it We observe up to $80\%$ reduction in the number of both \hising pulses and bit flips as compared to \unionofstars, while maintaining Max-Cut approximation $\ge 0.95$.}

Finally, we verify that our compilations reduce the amount of accumulated noise in two models. The first is a generic model for digital computations, where reducing the number of edges from $m$ to $m' < m$ leads to an exponentially larger lower bound on the circuit fidelity, $F'_0 = F_0^{m'/m}$. The second model considers the detailed physics of dephasing decoherence in trapped ions with global interactions, which is an important noise source in previous experiments with hundreds of trapped ions \cite{bohnet2016quantum}.  We use a Lindbladian master equation to describe the effect of the dephasing noise, ignoring other noise sources, and derive an exact analytic expression for the expected cost in QAOA.  The cost is exponentially suppressed with respect to the noiseless case, with an exponential prefactor that depends on the compilation time and a dephasing rate $\Gamma$.  We analyze QAOA performance with varying $\Gamma$ for our MQLib instances and compare the expected cost in the original compilation, with decomposition, and with decomposition and sparsification. We find decomposition has very significant benefits on maintaining a high-cost value in the presence of noise; sparsification is ineffective for this specific noise model, but we argue it would play an important role for noise sources that are not included in the present analytic treatment. We conclude our advanced compilation techniques are expected to provide significant benefits in reducing noise on quantum computing hardware, thus bridging some of the gap between classical computing and current quantum hardware for solving these problems.

We introduce some notation and give background on Max-Cut, QAOA, and \unionofstars\ in Section \ref{sec: preliminaries}. We give our theoretical improvements for the number of \hising pulses and bit flips in Section \ref{sec: weighted-graphs-faster-union-of-stars} and corresponding experiments in \ref{sec: simulations}. Section \ref{sec: noise-model} discusses our noise model for QAOA and corresponding results. We conclude in Section \ref{sec: conclusion}. Omitted proofs and additional experiments are deferred to the appendices.

\section{Preliminaries}\label{sec: preliminaries}

We discuss the Max-Cut problem first and give background for the \unionofstars\  compilation. Next, we briefly review the background on classical sparsification algorithms.

\paragraph{Graphs and Max-Cut.} For positive integer $n$, we denote $[n] = \{1, \ldots, n\}$ and $[0, n] = \{0, 1, \ldots, n\}$. 
We assume familiarity with basic definitions in graph theory; the interested reader is referred to \cite{west_introduction_2001} for a more detailed introduction.
Graphs will be denoted by letters $G, H$, and are always simple (i.e., have no loops or multiple edges) and undirected.
They may be weighted or unweighted.
An unweighted graph $G = (V, E)$ is specified by the set of vertices $V$ and edges $E$.
We will assume that the vertex set $V = [n] := \{1, \ldots, n\}$ for an arbitrary but fixed positive integer $n$. The number of edges $|E|$ is denoted by $m$.
All graphs are assumed to be connected so that $m \ge n - 1$. Further, $m \le \frac{n(n - 1)}{2} \le \frac{n^2}{2}$ since all graphs are assumed to be simple.
A weighted graph $G = (V, E, c)$ additionally has edge costs or weights $c: E \to \R_+$.
An unweighted graph $G = (V, E)$ is equivalent to the weighted graph $(V, E, c)$ with $c_e = 1$ for all $e \in E$.
We assume that $c \ge 0$, i.e., all edge weights are nonnegative.

The \emph{sum} of two graphs $G_1 = (V, E_1, c_1)$ and $G_2 = (V, E_2, c_2)$ is $G = (V, E_1 \cup E_2, c_1 + c_2)$.
In particular, for an unweighted graph $G = (V, E)$ and a partition $(E_1, \dots, E_T)$ of $E$, we have $G = \sum_{i \in [T]} G_i$ where $G_i = (V, E_i)$.
For a constant $\alpha \in \R$, graph $\alpha G$ has edge costs multiplied with $\alpha$, i.e., $\alpha G = (V, E, \alpha c)$.
For example, in Fig. \ref{fig: decomposition-example}, we have $G = G_1 + 2 G_2$.

Star graphs play a special role in the \unionofstars\  compilation of \cite{rajakumar_generating_2022}. A star graph $G = (V, E)$ is an unweighted graph with a special vertex $v^*$ called the central vertex and a subset $U \subseteq V \setminus \{v^*\}$ to which it is connected. That is, the only edges in $G$ are of the form $v^*u$ for $u \in U$.
Any unweighted graph $G = (V, E)$ can be written as the sum of (at most) $n - 1$ star graphs; we omit the proof of this simple lemma:
\begin{lemma}\label{lem: star-decomposition}
    {\it Any unweighted graph $G = (V, E)$ can be expressed as a sum $\sum_{i \in [T]} G_i$ where each $G_i$ is a star graph and $T \le n - 1$.} 
\end{lemma}

Given a graph $G = (V, E, c)$ and a vertex set $S \subseteq V$, the cut $\Delta_G(S)$ is the set of edges with exactly one endpoint in $S$, i.e., $\Delta_G(S) = \{uv \in E: |\{u, v\} \cap S| = 1\}$. The corresponding \emph{cut value} is the sum of costs of edges in the cut, and is denoted as $\delta_G(S) = \sum_{e \in \Delta_G(S)} c_e$. For brevity, when $G$ is clear from context, we denote this as $\delta(S)$. We often speak of `cut $S$' when we mean `cut $\Delta_G(S)$'.
The Max-Cut in $G$ is the cut corresponding to set ${\arg\max}_{S \subseteq V} \delta_G(S)$; the corresponding cut value of the maximum cut is $ \delta_G(S)$ and we often denote it by $\delta_G^{\max}$.

Given $G$, computing $\maxcut_G$ is NP-hard, i.e., there is no polynomial-time algorithm that given $G$ computes $\maxcut_G$ if P $\neq$ NP.
For $\alpha \in [0, 1]$, vertex set $S$ is called an $\alpha$-approximation for Max-Cut in $G$ if $\delta_G(S) \ge \alpha \cdot \maxcut_G$, where $\alpha = 1$ corresponds to the optimal cut. The celebrated Goemans-Williamson algorithm \cite{goemans_879-approximation_1994} is the classical state-of-the-art and gives a $\alpha^*$-approximation for any graph $G$ (in polynomial-time), where $\alpha^* \simeq 0.878$. Assuming the Unique Games Conjecture, no polynomial-time algorithm can achieve a better approximation ratio for all graphs \cite{khot_power_2002}. Assuming P $\neq$ NP, no polynomial-time algorithm can achieve approximation ratio $\ge 0.942$ for all graphs \cite{hastad_optimal_2001}.

We refer to a cut with value at most half the total sum of all edge costs (i.e., $\frac{1}{2} \sum_{e \in E} c_e$) as \emph{trivial}, and a cut with value $> \frac{1}{2} \sum_{e \in E} c_e$ as \emph{non-trivial}.
It is straightforward to obtain a cut with value $\frac{1}{2} \sum_{e \in E} c_e$, for example, using the greedy algorithm that iteratively adds or removes vertices from a cut $S$ until its cut value can no longer be improved. Since $\sum_{e \in E} c_e$ is an upper bound on $\delta^{\max}_G$, this shows that obtaining a $0.5$-approximation is straightforward.

Throughout, we reduce the Max-Cut problem on graph $G = (V, E, c)$ to a different graph $G' = (V, E', c')$ as follows: $G'$ is constructed so that the cut value $\delta_G(S)$ of any non-trivial cut $S \subseteq V$ in $G$ is $(1 - \epsilon)$ within the cut value $\delta_{G'}(S)$ of $S$ in $G'$. Therefore, solving Max-Cut in $G'$ up to an $\alpha$-approximation solves Max-Cut in $G'$ up to an $\alpha(1 - \epsilon)$-approximation in $G$, for any $\alpha \ge \frac{1}{2}$. In particular, obtaining the Max-Cut in $G'$ gives a $(1 - \epsilon)$-approximate cut in $G$.

Given graph $G = (V, E, c)$ with nonnegative edge weights, we use the QAOA variant that uses gates of the form $\exp(-i\gamma C)$ for the cost Hamiltonian $C = \sum_{uv \in E} c_{uv} \sigma_u^z \sigma_v^z$. For a vertex set $S \subseteq V$ and corresponding bit string $\bm z_S$, the expected values of $\langle C \rangle := \langle \bm z_S |C|\bm z_S \rangle$ and the cut value $\delta_G(S)$ are related as follows:
\begin{align}\label{eqn: operator-expectation-vs-max-cut-values}
    \langle C \rangle &:= \langle \bm z_S |C|\bm z_S \rangle = \sum_{\substack{uv \in E: \\ |\{u, v\} \cap S| \neq 1}} c_{uv} - \sum_{\substack{uv \in E: \\ |\{u, v\} \cap S| = 1}} c_{uv} \notag \\
    &= \sum_{uv \in E} c_{uv} - 2 \delta_G(S). 
\end{align}
Therefore, the minimum eigenvector of $C$ corresponds to the Max-Cut in $G$, which can be formulated as finding the maximum eigenvector of $\left(\frac{1}{2}\sum_{e \in E} c_{e}\right) I - \frac{1}{2}C$ \cite{farhi_quantum_2014}. 

Further, the lower $\langle \bm z_S| C | \bm z_S \rangle$ is, the larger the cut value $\delta_G(S)$. Since a non-trivial cut $S \subseteq V$ has cut value $\delta_G(S) > \frac{1}{2} \sum_{uv} c_{uv}$, we have that $\langle \bm z_S | C | \bm z_S\rangle < 0$ for non-trivial cuts $S$. 
This observation will be useful in Section \ref{sec: noise-model}, where we will compare a bit string $\bm z$ (generated using our algorithms) against baseline bit string $\bm z'$ using QAOA under noise in terms of their operator expectations $\langle \bm z|C| \bm z\rangle$, $\langle \bm z'|C| \bm z'\rangle$ to gauge their quality for Max-Cut.

\paragraph{Overview of \unionofstars.}

We next provide an overview of the \unionofstars\ algorithm \cite{rajakumar_generating_2022} for generating cost Hamiltonians $C$ for Max-Cut on trapped-ion devices using global \hising operations and bit flips.
One approach to \emph{compiling} these cost Hamiltonians $C = \sum_{uv \in E} c_{uv} \sigma_u^z \sigma_v^z$ on trapped-ion hardware is to start from the empty Hamiltonian $C = 0$ and apply a sequence of \hising pulses and bit flip operations. Each \hising pulse is implemented using a global MS operation and a pulse of length or time $t_p > 0$ adds all-to-all interaction terms $\sum_{u, v \in V} t_p \sigma_u^z \sigma_v^z$. The pulse may also be augmented with Pauli-$X$ "bit flips" to change the sign of $\sigma_v^z$ using $\sigma^x_v\sigma_v^z\sigma^x_v=-\sigma^z_v$. Using these ``flips'' on a set of vertices $S \subseteq V$ we obtain the general update rule for $C$: 
\begin{equation}\label{eqn: ising-update}
    C \gets C \pm \sum_{u, v \in V} (-1)^{\mathbf{1}_S(u) + \mathbf{1}_S(v)} t_p \sigma_u^z \sigma_v^z,
\end{equation}
where $\mathbf{1}_S(u)$ is the indicator for whether $u \in S$ and equals $1$ if $u \in S$ and is $0$ otherwise, while the sign $\pm$ of the update is a choice that can be made experimentally by changing the sign of the frequency of the laser driving the transition, as described in \cite{rajakumar_generating_2022}. Formally, we define graph compilation as follows:

\begin{definition}[Graph compilation]
    {\it A sequence of \hising pulses and bit flip operations that produces graph coupling (as described above) for a graph $G$ is called a graph compilation of $G$. $\nHising(G)$ is the minimum number of \hising pulses and $\nTotal(G)$ is the minimum number of total operations (\hising pulses and bit flips), in any graph compilation for $G$.}
\end{definition}

We say that the total time of the compilation for graph $G$ is the sum of the lengths or times $t_p$ of \hising pulses in the compilation\footnote{We omit the time taken by bit flip operations since they are at least an order of magnitude smaller in comparison to \hising pulses \cite{rajakumar_generating_2022}.}:
\begin{equation}\label{eqn: time-of-compilation}
    T = \sum_{p} t_p 
\end{equation}

Rajakumar \emph{et al.} \cite{rajakumar_generating_2022} conjecture that the problem of determining the shortest graph compilation for a given graph $G$ is NP-hard, i.e., there may be no polynomial-time algorithm to find the compilation with the fewest number of \hising pulses. However, they give a polynomial-time algorithm to obtain a (possibly sub-optimal) compilation with the number of \hising pulses upper bounded linearly in the input size: at most $3n - 2$ for unweighted graphs and at most $3m + 1$ for weighted graphs.

They start by noting that the graph $G_1 + G_2$ can be compiled by combining the compilations for each $G_i$ ($i=1, 2$) first:
\begin{lemma}[\cite{rajakumar_generating_2022}, Lemma 1]\label{lem: union-of-stars-graph-combination}
    Suppose $G = \alpha_1 G_1 + \alpha_2 G_2$ for weighted graphs $G_1, G_2$ on vertex set $V$ and reals $\alpha_1, \alpha_2 \in \R$. Then, if there exist graph compilations with $s_i$ \hising pulses and $t_i$ bit flips for graph $G_i$, $i \in \{1, 2\}$, then there exists a graph compilation for $G$ with $s_1 + s_2$ \hising pulses and $t_1 + t_2$ bit flips. In particular, $\nHising(G) \le \nHising(G_1) + \nHising(G_2)$ and $\nTotal(G) \le \nTotal(G_1) + \nTotal(G_2)$.
\end{lemma}

They also show the existence of compilations for all graphs and give a polynomial-time algorithm called \unionofstars\ that gives a specific sequence of \hising pulses and bit flips for a given graph.
Their construction observes that an unweighted star graph can be compiled using $4$ pulses.
By Lemma \ref{lem: star-decomposition}, this implies that an unweighted graph on $n$ vertices can be compiled using at most $4(n - 1)$ pulses, or $\nHising(G) \le 4(n - 1)$.
Further, since any single edge is also a star graph, they decompose any weighted graph with $m$ edges into its edges, and obtain $\nTotal(G) \le 4m$.
By combining redundant pulses, they improve these numbers to $3(n - 1) + 1$ and $3m + 1$ respectively:
\begin{theorem}[\cite{rajakumar_generating_2022}, Theorems 5, 6]\label{thm: undirected-graph-gc}
    $\nHising(G) \le 3n - 2$ for an unweighted graph and $\nHising(G) \le 3m + 1$ for a weighted graph.
\end{theorem}

The following result for $\nTotal$ is implicit in their construction:

\begin{lemma}\label{lem: gc_2_lemma}
    For any graph $G$ with $m$ edges, $\nTotal(G) = O(m)$.
\end{lemma}

\begin{algorithm*}
\hrule
\caption{\textsc{graph-sparsification-using-effective-resistances} \cite{spielman_graph_2011}}
\label{alg: sparsification}
\begin{algorithmic}[1]
\Require weighted graph $G = (V, E, c)$ and parameter $\epsilon > 0$
\Ensure sparsifier $H = (V, E', c')$ of $G$
\State compute effective resistances $r_e$ for each $e \in E$
\State set $q = \frac{C_0 n \log n}{\epsilon^2}$ for a large enough constant $C_0$
\For{$t$ = 1 to $q$, independently}:
    \State choose a random edge $e \in E$ with probability $p_e$ proportional to $c_e r_e$
    \State add $e$ to $H$ with weight $\frac{c_e}{qp_e}$, summing weights if $e$ is already in $H$
\EndFor
\State \textbf{return} $H$
\end{algorithmic}
\hrule
\end{algorithm*}

\paragraph{Graph sparsification.} The theory of graph sparsifiers was initiated by Karger in 1994  \cite{karger_random_1994}.
Given any input graph $G = (V, E, c)$ on $n$ vertices and an error parameter $\epsilon > 0$, graph $H = (V, E', c')$ is called a \emph{graph} sparsifier of $G$ if (a) $G, H$ have similar cut values, i.e., for each $S \subseteq V$, the cut value $\delta_G(S)$ is within factor $1 \pm \epsilon$ of the cut value $\delta_H(S)$ and (b) $H$ has at most $O\left(n \log^\kappa n/\epsilon^2\right)$ edges for some constant $\kappa$. Note that $\delta_H(S)$ is computed with respect to the edge cost function $c'$ of $H$.

The current best provable graph sparsification algorithms can reduce the number of edges to almost linear in the vertices, with some loss in approximation for the cut values of the graph, \emph{for all cuts in the graph}:
\begin{theorem}[\cite{batson_twice-ramanujan_2014}, Theorem 1.1]\label{thm: sparsification}
    There exists a polynomial-time algorithm that given a weighted graph $G = (V, E, c)$ on $n$ vertices and error parameter $\epsilon > 0$, finds another weighted graph $H = (V, E', c')$, such that with high probability\endnote{We borrow this notation from algorithmic graph theory and say an event occurs `with high probability' if the corresponding probability $p_n$ goes to $1$ as the number $n$ of vertices of the graph goes to infinity, i.e., $p_n = 1 - o(1)$.} 
    \begin{enumerate}[label=(\roman*)]
        \item All cut values are preserved, i.e., $1 - \epsilon \le \frac{\delta_{H}(S)}{\delta_G(S)} \le 1 + \epsilon, ~~\forall S \subseteq V$,
        \item $H$ is sparse, i.e.,
        $|E'| = O\left(\frac{n}{\epsilon^2}\right)$.
    \end{enumerate}
\end{theorem}

In other words, if the Max-Cut was found on the sparse graph $H$, which can be computed classically, then it would approximate the Max-Cut on $G$ with $(1-\epsilon)$-approximation. Note that this result approximates {\it all} the cuts in the graph\footnote{There is another stream of literature that focuses on sparsifying the set of vertices in the graph, so that the minimum cuts across a given smaller set of terminals is preserved \cite{moitra2009approximation, charikar2010vertex}. However, it is unclear how this can be used to obtain guarantees for approximating the Max-Cut.}. The algorithm of \cite{batson_twice-ramanujan_2014} is significantly involved, and therefore, for the (classical) experiments in this work, we will instead use the simpler and more efficient sparsification algorithm of \cite{spielman_graph_2011} based on computing effective resistances\endnote{Effective resistance of an edge in a graph is the equivalent resistance measured between its endpoints, treating the edge weights as inverse resistances. For details on computing this quantity, we refer the interested reader to \cite{spielman_graph_2011}.}. This algorithm outputs sparsifiers with a slightly weaker guarantee of $|E'| = O\left(\frac{n \log n}{\epsilon^2}\right)$ for the number of edges. We state it in Algorithm \customnameref{alg: sparsification} for completeness (but without proof). We will use these sparsification results as black boxes and modify the {\unionofstars} compilation to provide guarantees on $\nHising$ and $\nTotal$ accordingly.

\section{Trapped-ion compilations using sparse \unionofstars}\label{sec: weighted-graphs-faster-union-of-stars}

In Section \ref{sec: faster-gc-1}, we present two variants of our decomposition algorithm and reduce the number of \hising pulses required in \unionofstars\ (i.e. $\nHising$) for weighted graphs from $O(m)$ to $O(n \log(n/\epsilon))$. Both decomposition algorithms use the facts that (1) removing edges with very small edge weight does not change the cut value of large cuts, and (2) the remaining graph can be written as the weighted sum of a small number of unweighted graphs.
In Section \ref{sec: second-graph-coupling-number}, we improve the total number of gates (i.e., \hising pulses and bit flips or $\nTotal$) for all graphs from $O(m)$ to $O\left(\frac{n \log(n/\epsilon)}{\epsilon^2}\right)$. We do this by first decomposing the graph to reduce $\nHising$, and the applying sparsification to reduce the number of edges in the decomposed graph. In Section \ref{sec: simulations}, we run experiments on the MQLib graph library and show that our algorithms significantly outperform \unionofstars\ in classical simulations (assuming no noise).

\subsection{Reduction in number of \hising pulses for weighted graphs}\label{sec: faster-gc-1}

First, we note that improving $\nHising(G)$ from $O(m)$ to $O\left(\frac{n}{\epsilon^2}\right)$ is an immediate consequence of sparsification (Theorem \ref{thm: sparsification}) and the \unionofstars\ upper bound (Theorem \ref{thm: undirected-graph-gc}). In this section, we give two algorithms to improve the dependence of $\nHising(G)$ on $\epsilon$, 
\begin{enumerate}
    \item Algorithm \customnameref{alg: exponential-decomposition}, which improves it to $O(\frac{n}{\epsilon} \log\left(\frac{n}{\epsilon}\right))$ and
    \item Algorithm \customnameref{alg: binary-decomposition}, which further improves it to $O\left(n \log\left(\frac{n}{\epsilon}\right)\right)$.
\end{enumerate}
This latter bound is logarithmic in $1/\epsilon$ and is therefore an \emph{exponential} improvement over $O\left(\frac{n}{\epsilon^2}\right)$.

Both algorithms work as follows: first, we decompose the given weighted graph $G = (V, E, c)$ into unweighted graphs $G_1, \ldots, G_k$ and weights $\alpha_1, \ldots, \alpha_k > 0$ such that $G \simeq \sum_{j \in [k]} \alpha_j G_j$ (we shortly define $\simeq$). Since each $G_j$ is unweighted, it takes at most $O(n)$ \hising pulses to compile using \unionofstars\ (see Theorem \ref{thm: undirected-graph-gc}). Then, we combine these compilations using Lemma \ref{lem: gc_2_lemma} into a single compilation for $G$ that takes $O(kn)$ \hising pulses. For Algorithm \customnameref{alg: exponential-decomposition}, $k = O\left(\frac{1}{\epsilon} \log\left(\frac{n}{\epsilon}\right)\right)$ while Algorithm \customnameref{alg: binary-decomposition} improves this to $k = O\left(\log \frac{n}{\epsilon}\right)$. We begin with the description and formal guarantee for Algorithm \customnameref{alg: exponential-decomposition}, followed by Algorithm \customnameref{alg: binary-decomposition}. All proofs are deferred to Section \ref{sec: algorithms-analysis}.

\begin{algorithm}[t]
\hrule
\caption{\textsc{exp-decompose}}\label{alg: exponential-decomposition}
\begin{algorithmic}[1]
\Require weighted graph $G = (V, E, c)$ and parameter $\epsilon > 0$
\Ensure weighted graph $G'$ such that $1 - \epsilon \le \frac{\maxcut_{G'}}{\maxcut_{G}} \le 1$
\State set $c^* = \max_{e \in E} c(e)$
\State set $\tau = \frac{\epsilon c^*}{2 n^2}$ and set $k = \lceil \log_{1 + \epsilon/2} (c^*/\tau) \rceil$
\For{$j \in [0, k]$}
    \State initialize undirected graph $G_j = (V, E_j)$ with $E_j = \emptyset$
\EndFor
\For{each edge $e \in E$}
    \If{$c(e) > \tau$}
        \State let $j$ be the unique integer in $[k]$ such that $\tau (1 + \epsilon/2)^{j - 1} < c(e) \le \tau(1 + \epsilon/2)^{j}$ \label{step: exp-decompose-edge-weight-multiplicative-approximation}
        \State add edge $e$ to $E_j$
    \EndIf
\EndFor
\State \textbf{return} weighted graph $G' = (V, E', c')$ defined as
\[
    G' = \sum_{j \in [k]} (1 + \epsilon/2)^{j - 1} G_j.
\]
\end{algorithmic}
\hrule
\end{algorithm}

\paragraph{Algorithm \customnameref{alg: exponential-decomposition}.} Intuitively, the algorithm obtains graphs $G_1, \ldots, G_k$ by grouping edges in $G$ that have similar weights (within a factor $\simeq (1 + \epsilon/2)$). Crucially, edges with very small costs are removed entirely, as follows: denote the max cost as $c^* = {\max}_{e \in E} c(e)$, then all edges with costs smaller than the \emph{threshold} $\tau := \frac{\epsilon c^*}{2n^2}$ are removed. Since the total cost of such edges is at most $m \times \tau \le \frac{n^2}{2} \times \frac{\epsilon c^*}{2n^2} = \frac{\epsilon}{4} c^* \le \frac{\epsilon}{4} \maxcut_G$, this reduces the Max-Cut in $G$ by factor at most $(1 - \epsilon/4)$. The details are presented in Algorithm \customnameref{alg: exponential-decomposition}. We present the theoretical guarantee of the algorithm next:

\begin{theorem}\label{thm: exp-decompose}
    Given a weighted graph $G = (V, E, c)$ and error parameter $\epsilon \in (0, 1)$, \customnameref{alg: exponential-decomposition} returns another weighted graph $G' = (V, E', c')$ such that we get
    \begin{enumerate}[label=(\roman*)]
        \item Preservation of non-trivial cuts, i.e., for all $S \subseteq V$ with $\delta_S(G) \ge \frac{1}{2} \sum_{e \in E} c_e$,
        \[
            1 - \epsilon \le \frac{\delta_{G'}(S)}{\delta_{G}(S)} \le 1.
        \]
        In particular, the Max-Cut in $G'$ is a $(1 - \epsilon)$-approximate cut in $G$.
        \item Reduction in \hising pulses: $\nHising(G') = O\left(\frac{n}{\epsilon} \log\frac{n}{\epsilon}\right)$.
    \end{enumerate}
    Further, the number of edges in $G'$ is at most the number of edges in $G$.
\end{theorem}

\begin{algorithm}[h]
\hrule
\caption{\textsc{binary-decompose}}\label{alg: binary-decomposition}
\begin{algorithmic}[1]
\Require weighted graph $G = (V, E, c)$ and parameter $\epsilon > 0$
\Ensure weighted graph $G'$ such that $1 - \epsilon \le \frac{\maxcut_{G'}}{\maxcut_G} \le 1$
\State set $c^* = \max_{e \in E} c(e)$ and $\eta = \frac{\epsilon c^*}{n^2}$
\State set $k = 1 + \lfloor \log_2(n^2/\epsilon) \rfloor$
\For{each edge $e \in E$}
    \State set $d(e) = \lfloor \frac{c(e)}{\eta} \rfloor$
    \State write $d(e) \in [0, n^2/\epsilon]$ in binary with $k$ digits
    \[
        b_{k - 1}(e) b_{k - 2}(e) \ldots b_0(e)
    \]
\EndFor
\For{$j \in [k]$}
    \State let $G_j = (V, E_j)$ be the graph such that $e \in E_j$ iff $b_{j - 1}(e) = 1$ 
\EndFor
\State \textbf{return} weighted graph $G' = (V, E', c')$ defined as
\[
    G' = \eta 2^{k - 1} G_k + \eta 2^{k - 2} G_{k - 1} + \ldots + \eta 2^0 G_1.
\]
\end{algorithmic}
\hrule
\end{algorithm}

\paragraph{Algorithm \customnameref{alg: binary-decomposition}.} Our second algorithm decomposes $G \simeq \sum_{j \in [k]} \alpha_j G_j$ by first computing the binary representation of each edge cost $c(e)$ up to $k$ digits. The $j$th unweighted graph $G_j$ contains exactly those edges with the $j$th bit in their binary representation equal to $1$, and the weights $\alpha_j$ increase in powers of $2$. For error parameter $\epsilon > 0$, we choose $k = 1 + \lfloor \log_2 (n^2/\epsilon) \rfloor$ to ensure that Max-Cut in decomposition $\sum_{j \in [k]} \alpha_j G_j$ is within factor $(1 + \epsilon)$ of the Max-Cut in $G$ (see proof details in Appendix \ref{sec: algorithms-analysis}). The details are presented in Algorithm \customnameref{alg: binary-decomposition} and the next theorem contains the formal guarantee:

\begin{theorem}\label{thm: faster-union-of-stars-weighted-graphs}
    Given a weighted graph $G = (V, E, c)$ and error parameter $\epsilon \in (0, 1)$, \customnameref{alg: binary-decomposition} returns another weighted graph $G' = (V,E',c')$ such that we get
    \begin{enumerate}[label=(\roman*)]
        \item Preservation of non-trivial cuts, i.e., for all $S \subseteq V$ with $\delta_S(G) \ge \frac{1}{2} \sum_{e \in E} c_e$,
        \[
            1 - \epsilon \le \frac{\delta_{G'}(S)}{\delta_{G}(S)} \le 1.
        \]
        In particular, the Max-Cut in $G'$ is a $(1 - \epsilon)$-approximate cut in $G$.
        \item Reduction in \hising pulses: $\nHising(G') = O\left(n \log\frac{n}{\epsilon}\right)$.
    \end{enumerate}
    Further, the number of edges in $G'$ is at most the number of edges in $G$.
\end{theorem}

These algorithms have a significant impact on the reduction of the number of \hising pulses in the compilation. This will be evident from our classical experiments that simply count these operations for the modified compilation.

\subsection{Reduction in total number of operations}\label{sec: second-graph-coupling-number}

\begin{algorithm*}[t]
\hrule
\caption{\textsc{sparse-union-of-stars}}\label{alg: faster-union-of-stars-2}
\begin{algorithmic}[1]
\Require weighted graph $G = (V, E, c)$ and parameters $\epsilon_1, \epsilon_2 > 0$
\Ensure weighted graph $G = (V, E', c')$
\State $H =$ \textsc{graph-sparsification-using-effective-resistances}
$(G, \epsilon_1)$
\State $G'_{\text{bin}} = \customnameref{alg: binary-decomposition}(H, \epsilon_2)$ and $G'_{\text{exp}} = \customnameref{alg: exponential-decomposition}(H, \epsilon_2)$
\State set
    \begin{equation*}
        G' := \begin{cases}
            G'_{\text{bin}} & \mathrm{if} \ \nHising(G'_{\text{bin}}) < \nHising(G'_{\text{exp}}) \\
            G'_{\text{exp}} & \mathrm{if} \ \nHising(G'_{\text{bin}}) \ge \nHising(G'_{\text{exp}})
        \end{cases}
    \end{equation*}
\State \textbf{return} $G'$ and $ \unionofstars(G')$
\end{algorithmic}
\hrule
\end{algorithm*}

\customnameref{alg: binary-decomposition} and \customnameref{alg: exponential-decomposition} reduce $\nHising$ for weighted graphs, but they may not reduce the total number of operations which also involve bit flips. Recall that $\nTotal(G)$ is upper bounded by $O(m)$ for all graphs with $m$ edges. We combine our decomposition idea with sparsification to reduce the number of edges (and hence the $\nTotal$) while still maintaining $\nHising = O(n\log(n/\epsilon))$.

\begin{theorem}\label{thm: second-graph-coupling-number}
    Given a weighted graph $G = (V, E, c)$ and error parameter $\epsilon \in (0, 1)$, \customnameref{alg: faster-union-of-stars-2} with $\epsilon_1 = \epsilon_2 = \frac{\epsilon}{3}$ returns another weighted graph $G' = (V, E', c')$ such that with high probability, we get
    \begin{enumerate}[label=(\roman*)]
        \item Preservation of non-trivial cuts, i.e., for all $S \subseteq V$ with $\delta_S(G) \ge \frac{1}{2} \sum_{e \in E} c_e$,
        \[
            1 - \epsilon \le \frac{\delta_{G'}(S)}{\delta_{G}(S)} \le 1 + \epsilon.
        \]
        In particular, the Max-Cut in $G'$ is a $(1 - \epsilon)$-approximate cut in $G$.
        \item Reduction in \hising Pulses:   $\nHising(G') = O\left(n \log \frac{n}{\epsilon}\right)$,
        \item Reduction in total number of operations:  $\nTotal(G') = O\left(\frac{n \log(n/\epsilon)}{\epsilon^2}\right)$.
    \end{enumerate}  
\end{theorem}

Note that achieving a graph $G'$ with guarantee (i) and with $\nTotal = O(n/\epsilon^2)$ follows directly from Lemma \ref{lem: gc_2_lemma} and Theorem \ref{thm: sparsification}. However, having an additional guarantee on $\nHising$ requires the decomposition technique. The detailed proof is presented in the proof in Appendix \ref{sec: algorithms-analysis}.

\begin{figure*}
    \centering
    \includegraphics[width=0.6\linewidth]{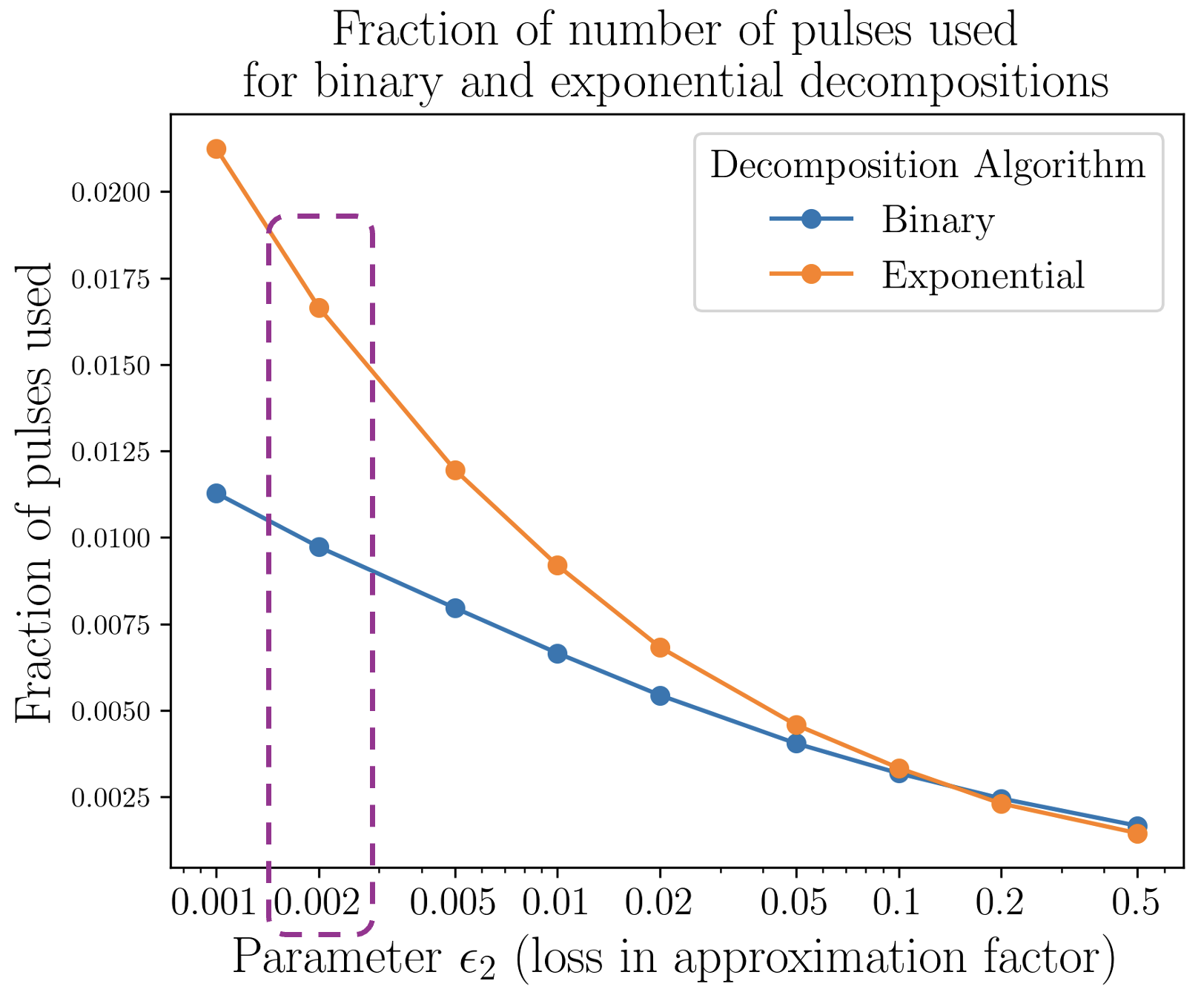}
    \caption{At a loss of $\epsilon_2$ in the approximation factor, the fraction of \hising pulses used by \customnameref{alg: binary-decomposition} and \customnameref{alg: exponential-decomposition} for the weighted graph \texttt{g001737} from MQLib with $n = 952$ vertices and $m = 430,730$ edges is depicted in blue and orange respectively. Compared to the edge-by-edge \unionofstars\ construction, \customnameref{alg: binary-decomposition} and \customnameref{alg: exponential-decomposition} use $1\%$ and $1.75\%$ pulses respectively, while maintaining a 0.998-approximation factor for Max-Cut for the original graph (corresponding to $\epsilon_2 = 0.002$, highlighted in purple). The gap between the performance of the algorithms reduces as the loss $\epsilon_2$ of the approximation factor is increased. This pattern holds for other large and dense graphs in MQLib, see Appendix \ref{app: bin-decompose} for more examples.}
    \label{fig: exp-vs-bin-decompose-large-graph}
\end{figure*}

\textbf{Choice of the decomposition algorithm.} Note that in Algorithm \customnameref{alg: faster-union-of-stars-2}, we choose \emph{the best of} \customnameref{alg: exponential-decomposition} and \customnameref{alg: binary-decomposition}, i.e., whichever requires a smaller number of \hising pulses. Since both these algorithms are classical, this is not computationally prohibitive for QAOA. This is useful since \customnameref{alg: exponential-decomposition} often performs better than its \emph{worst-case} theoretical guarantee from Theorem \ref{thm: exp-decompose} suggests, particularly for smaller graphs and larger values of error parameter $\epsilon_2$. We emphasize that this is not always the case and Figure \ref{fig: exp-vs-bin-decompose-large-graph} shows an example of a large graph ($\simeq 400,000$ edges) where \customnameref{alg: binary-decomposition} performs better. Appendix \ref{app: bin-decompose} discusses this in greater detail.

\subsection{Compilations using \customnameref{alg: faster-union-of-stars-2} on the MQLib graph library}\label{sec: simulations}

\begin{figure*}
    \begin{minipage}[t]{0.49\textwidth}
        \centering
        \includegraphics[width=\textwidth]{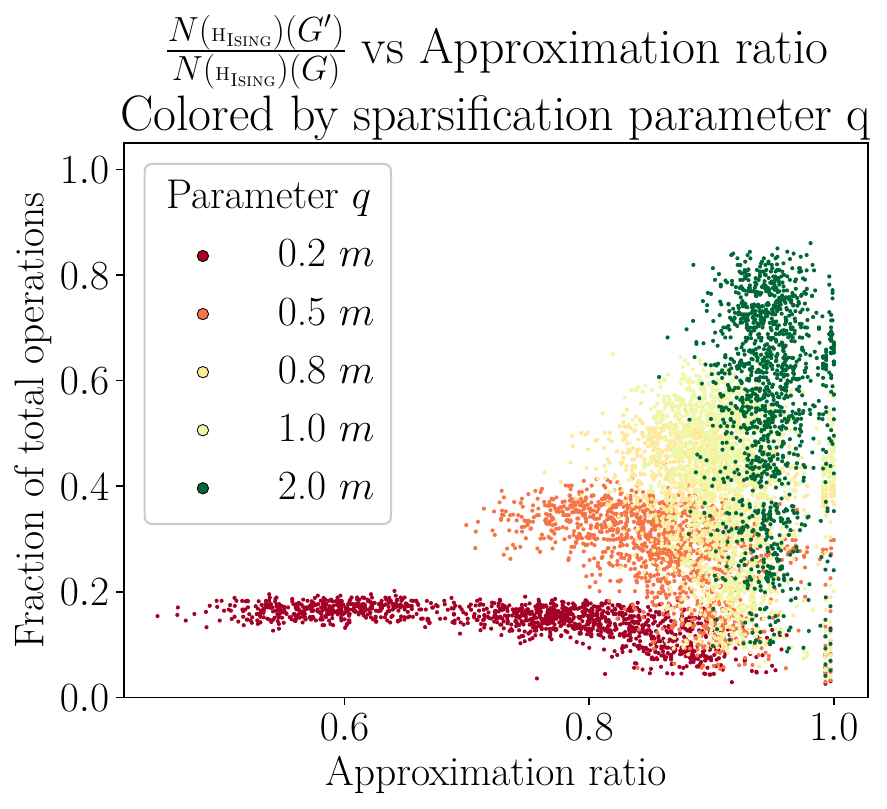}
        \caption{Reduction in $\nHising$ (total number of \hising pulses) vs Max-Cut approximation for various runs of our experiment on weighted graphs in MQLib, colored by parameter $q$. Each data point is a single run. Points on the lower right have high reductions and high Max-Cut approximation.}
        \label{fig: number-of-pulses-sparsification}
    \end{minipage}
    \hfill
    \begin{minipage}[t]{0.49\textwidth}
        \centering
        \includegraphics[width=\textwidth]{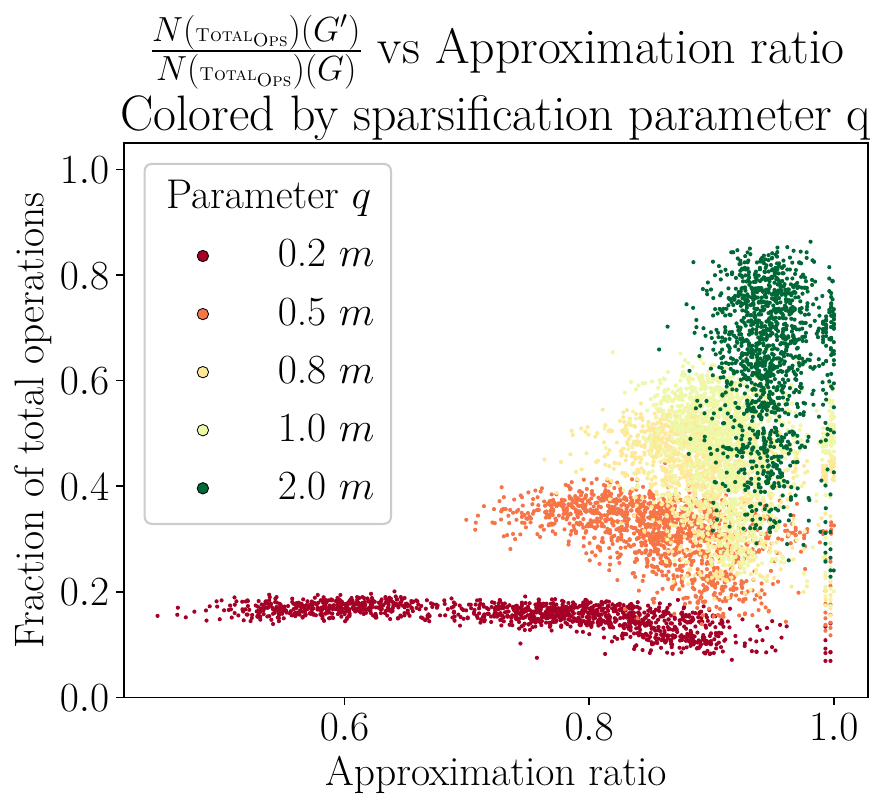}
        \caption{Reduction in $\nTotal$ (total number of \hising pulses and bit flip gates) vs Max-Cut approximation for various runs of our experiment on weighted graphs in MQLib. Each data point is a single run. Points on the lower right have high reductions and high Max-Cut approximation.}
        \label{fig: total-gates-sparsification}
    \end{minipage}
\end{figure*}

\begin{table*}[]
\caption{Average reduction in \hising pulses, total operations, and compilation time for parameter combinations of sparsification and decomposition that yield an average Max-Cut approximation ratio $\frac{\delta^{\max}_{G'}}{\delta^{\max}_G} \geq 0.90$ across graphs in the MQLib library (with $50 \le n < 200$ and $2n \le m < 2000$), using Algorithm~\customnameref{alg: faster-union-of-stars-2}. Here $G$ is the original input graph, and $G'$ is the decomposed + sparsified instance. Smaller numbers are better for $\nHising, \nTotal$, and compilation time columns. Numbers closer to $1$ are better for the ratio of the Max-Cut approximation column. Smaller $q$ gives more sparsified instances.}
\label{tab: average-reduction-mqlib}
\hspace*{-0.4cm}
\centering
\small
\begin{tabular}{|c|c||c|c|c||c|}
\hline
\begin{tabular}[c]
{@{}c@{}}Sparsification\\ parameter $q$\end{tabular} &
  \begin{tabular}[c]{@{}c@{}}Decomposition\\ parameter $\epsilon_2$\end{tabular} &
  $\dfrac{\nHising(G')}{\nHising(G)}$ &
  $\dfrac{\nTotal(G')}{\nTotal(G)}$ &
  $\dfrac{\substack{\text{\large compilation} \\ \text{\large time of }{G'}}}{\substack{\text{\large compilation} \\ \text{\large time of } G}}$ &
  \begin{tabular}[c]{@{}c@{}}Max-Cut \\ approximation \\ for $G$ attained\\ using the Max-Cut \\ in $G'$\end{tabular} \\ \hline
0.8 & 1.00 & 0.352 & 0.569 & 0.404 & 0.901 \\ \hline
1.0 & 0.10 & 0.535 & 0.870 & 0.557 & 0.915 \\ \hline
1.0 & 0.25 & 0.488 & 0.780 & 0.523 & 0.917 \\ \hline
1.0 & 0.50 & 0.443 & 0.681 & 0.491 & 0.914 \\ \hline
1.0 & 0.75 & 0.419 & 0.614 & 0.474 & 0.916 \\ \hline
1.0 & 1.00 & 0.400 & 0.567 & 0.460 & 0.916 \\ \hline
1.0 & 2.00 & 0.351 & 0.439 & 0.425 & 0.913 \\ \hline
1.0 & 5.00 & 0.301 & 0.299 & 0.389 & 0.912 \\ \hline
2.0 & 0.10 & 0.733 & 0.873 & 0.762 & 0.949 \\ \hline
2.0 & 0.25 & 0.668 & 0.774 & 0.716 & 0.949 \\ \hline
2.0 & 0.50 & 0.605 & 0.668 & 0.670 & 0.948 \\ \hline
2.0 & 0.75 & 0.561 & 0.594 & 0.638 & 0.946 \\ \hline
2.0 & 1.00 & 0.530 & 0.537 & 0.616 & 0.945 \\ \hline
2.0 & 2.00 & 0.454 & 0.406 & 0.562 & 0.947 \\ \hline
2.0 & 5.00 & 0.364 & 0.266 & 0.498 & 0.944 \\ \hline
\end{tabular}
\end{table*}

To exemplify the impact of sparsification and decomposition on the number of operations in graph compilation, we classically simulate the Algorithm \customnameref{alg: faster-union-of-stars-2} on graphs from the MQLib library \cite{dunning_what_2018}, a collection of challenging graphs for the Max-Cut problem designed to benchmark Max-Cut algorithms\footnote{The source code for computations in this section can be found at \url{https://github.com/jaimoondra/union-of-stars} \cite{moondra2026union_github}.}. To keep the computation of exact Max-Cut manageable, we choose all positive weighted graphs in MQLib with $50 \le n < 200$ and $2n \le m < 2000$. The condition $m \ge 2n$ excludes very sparse graphs (where edge-by-edge compilations are inexpensive anyway), resulting in 161 graphs in total. We refer to the original graph as $G$ and the corresponding modified instance as $G^\prime$. We compare vanilla \unionofstars\ and \customnameref{alg: faster-union-of-stars-2} for various combinations of sparsification and decomposition parameters on three metrics: (1) number of \hising pulses or $\nHising(H)$ value, (2) number of total gates or $\nTotal(H)$ value, and (3) total compilation time $T$ (see eqn. (\ref{eqn: time-of-compilation})), where $H \in \{G, G^\prime\}$. 

\paragraph{Choice of algorithm parameters.} We parameterize each run by $q$ that denotes the number of edges sampled from the original graph $G$ to get sparsified instance $G^\prime$ in Algorithm \customnameref{alg: sparsification}; recall that $q$ and error parameter $\epsilon_1$ are related as $q = \frac{C n \log n}{\epsilon_1^2}$ for a suitable constant $C$. For each graph $G$ with $m$ edges, we run \customnameref{alg: faster-union-of-stars-2} for various combinations of $\allowdisplaybreaks q \in \{0.2m, 0.5m, 0.8m, 1.0m, 2.0m\}$
and $\allowdisplaybreaks \epsilon_2 \in \{0.1, 0.25, 0.5, 0.75, 1.0, 2.0, 5.0\}$. 
Lower values of $q$ correspond to greater sparsification and higher values of $\epsilon_2$ correspond to stronger effect of decomposition. Note that while our theoretical guarantees for decomposition algorithms (Theorem \ref{thm: exp-decompose}, \ref{thm: faster-union-of-stars-weighted-graphs}) assume that $\epsilon_2 \in (0, 1)$, in practice higher values of $\epsilon_2$ perform very well while retaining a high approximation ratio.

\begin{figure*}
    \begin{minipage}[t]{0.49\textwidth}
        \centering
        \includegraphics[width=\columnwidth]{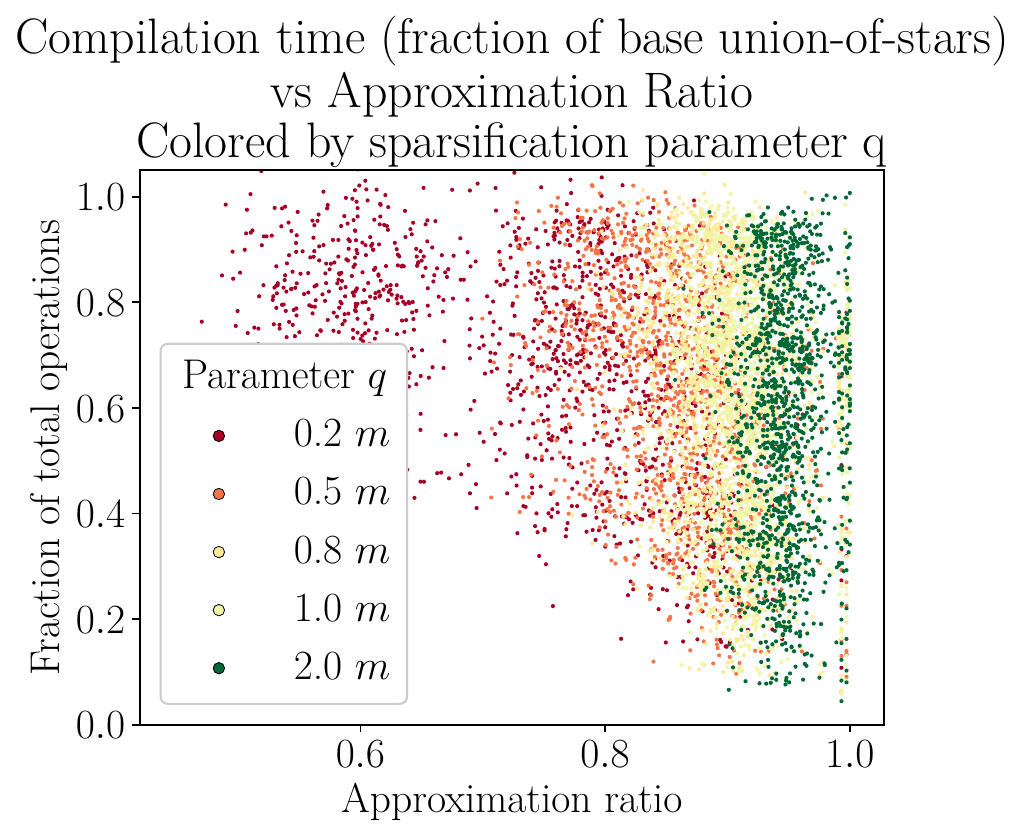}
        \caption{Reduction in the time $T = \sum_p t_p$ of compilation vs Max-Cut approximation for various runs of our experiment on weighted graphs in MQLib. Each data point is a single run.}
        \label{fig: time-of-pulses-sparsification}
    \end{minipage}
    \hfill
    \begin{minipage}[t]{0.49\textwidth}
        \centering
        \includegraphics[width=\columnwidth]{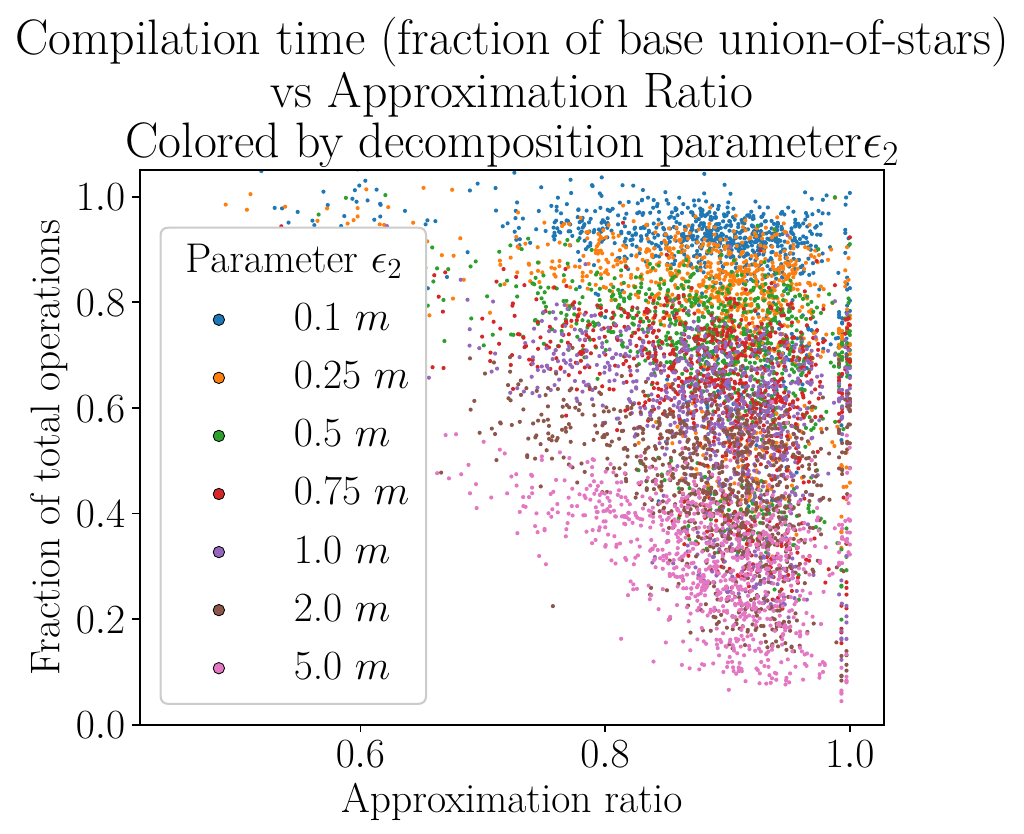}
        \caption{Reduction in the time $T = \sum_p t_p$ of compilation vs Max-Cut approximation for various runs of our experiment on weighted graphs in MQLib. Each data point is a single run.}
        \label{fig: time-of-pulses-decomposition}
    \end{minipage}
\end{figure*}

\paragraph{Results.} 

Fig. \ref{fig: number-of-pulses-sparsification} shows the fractional reduction $\frac{\nHising(G')}{\nHising(G)}$ using \customnameref{alg: faster-union-of-stars-2} as compared to \unionofstars, plotted against the Max-Cut approximation of $G'$ for $G$. Each data point represents one run of the experiment on a specific graph and with specific values of parameters $q$ and $\epsilon_2$.
For most graphs $G$, we observe over $40\%$ reduction in $\nHising$ value, while still recovering over $90\%$ of the Max-Cut in the original graph for suitable choices of the parameters (e.g., points in purple and red).
Further, there is a clear trade-off in the Max-Cut approximation and the fractional reduction $\frac{\nHising(G')}{\nHising(G)}$; higher reduction necessarily reduces the quality of the approximation. There is also a clear dependence on $q$: sparser graphs lead to a greater reduction in $\nHising$ value but obtain poorer approximation. Similar results hold for the total number of operations (Fig. \ref{fig: total-gates-sparsification}).

Similar reductions also hold for the total time of pulses. This is shown in Figs. \ref{fig: time-of-pulses-sparsification} and \ref{fig: time-of-pulses-decomposition}; the figures are identical except they are colored by the sparsification parameter $q$ and the decomposition parameter $\epsilon_2$, respectively. In Fig. \ref{fig: time-of-pulses-decomposition} observe that decomposition has a significant impact in the reduction of the total time of pulses, while sparsification (Fig. \ref{fig: time-of-pulses-sparsification}) does not have any significant correlation. We conclude from Figs.~\ref{fig: number-of-pulses-sparsification}-\ref{fig: time-of-pulses-decomposition} that sparsification reduces the total number of operations, while decomposition reduces the total time of pulses.

The total compilation time $T$ does not depend on the sparsification parameter $q$ because, in edge-by-edge compilations, $T$ is determined by the total weight of the edges in the graph--which remains largely unaltered by sparsification. As an example, consider a complete bipartite graph $G$ on vertex set $A \cup B$ with $|A| = |B| = n/2$ with edge set $E(G) = \{uv: u \in A, v \in B\}$ with each edge weight $ = 1$. Then it can be shown that the following (weighted) graph $H$ on vertex set $A \cup B$ is a sparsifier of $G$: for each $u \in A, v \in B$, include edge $uv \in E(H)$ with probability $p = \frac{1}{\sqrt{n}}$ independent of other edges. Assign weight $\frac{1}{p} = \sqrt{n}$ to this edge. In an edge-by-edge construction, the total length of pulses in $G$ is the sum of edge weights (up to constants), which is $\simeq n^2/4$, the same as for $H$. However, the number of pulses for $G$ is $O(|E(G)|) = O(n^2)$ while the number of pulses for $H$ is $O(|E(H)|) = O(p|E(G)|) = O(n^{3/2})$.

\section{Quantum noise analysis}\label{sec: noise-model}

Here, we consider the expected decrease in noise during quantum computations that utilize graph sparsification and decomposition techniques for compiling MaxCut with QAOA.  We first present a device-agnostic theoretical analysis for compiling sparsified instances with digital quantum computers using local one- and two-qubit gates, then analyze the expected noise in analog quantum computations with decomposition and sparsification.  

\subsection{Sparsified compilations of QAOA for digital quantum computations} \label{digital section}

Digital quantum computations are compiled to hardware-native universal gate sets that act on one or two qubits at a time. This applies to superconducting qubits \cite{bravyi2022future}, certain rydberg atom devices \cite{bluvstein2024logical}, trapped ions implementing two-qubit gates \cite{pino2021demonstration}, and any other digital quantum computing technology. For concreteness we consider a hardware gate set containing controlled-not $\mathrm{CNOT}_{uv} = U_{uv} $ operators as well as generic single-qubit rotations $\exp(-i \theta \sigma^\alpha_v) = U_v^\alpha(\theta)$ where $\alpha \in \{x,y,z\}$, though similar considerations apply to other hardware gates sets. We do not consider specific hardware connectivities or SWAP gate requirements, but instead assume that each qubit may interact with each other qubit, to keep the discussion generic. 

The QAOA unitary operator $\exp(-i \gamma C) = \prod_{(u,v) \in E} \exp(-i \gamma \sigma^z_u \sigma^z_v)$ is compiled from component operations $\exp(-i \gamma \sigma^z_u \sigma^z_v)$ for each edge in the graph. Each edge operation is compiled into our native gate set as
\begin{equation}\exp(-i \gamma \sigma^z_u \sigma^z_v) = U_{uv} U^z_v(\gamma) U_{uv}. \end{equation} 
For a quantum state described by a density operator $\rho$, the ideal evolution under any operator $U$ is
\begin{equation} \label{rho ideal}\rho' = U\rho U^\dag \end{equation}
where $\dag$ denotes the Hermitian conjugate. Applying component gates in sequence we obtain the desired unitary dynamics under the coupling operator unitary $\rho \leftarrow e^{-i \gamma C}\rho e^{i\gamma C}$.

Noisy quantum circuit operations can be described in the quantum channel formalism using Krauss operators $\sqrt{p_k}E_k$ with $\sum_k p_k E_k^\dag E_k=\mathbb{1}$, with $\mathbb{1}$ the identity operator, $p_k \geq 0$, and $\sum_k p_k = 1$ \cite{nielsen2001quantum}. Analogous to (\ref{rho ideal}) this gives
\begin{equation} \label{rho noise} \rho' = \sum_k p_kE_k U \rho U^\dag E_k^\dag, \end{equation} 
where we have used a notation in which the intended dynamics under $U$ is separated from the $E_k$.  Equation (\ref{rho noise}) is mathematically equivalent to a probabilistic model where evolution $E_kU$ is generated with probability $p_k$.  We consider $E_1 = \mathbb{1}$ as the ideal evolution, with probability $p_1$, while $E_k \neq \mathbb{1}$ for $k \geq 2$ represent various types of noisy evolution with probabilities $p_k$.  Applying a sequence of gates $U^{(i)}$ to compile the graph coupling operator $\exp(-i \gamma C) = \prod_{i=1}^{N_\mathrm{gates}} U^{(i)}$, the final state after these circuit operations can be expressed as 
\begin{equation} \rho_\mathrm{final} = F_0\rho_\mathrm{ideal} + (1-F_0)\rho_\mathrm{noise} \end{equation} 
where 
\begin{equation} F_0 = \prod_{i=1}^{N_\mathrm{gates}} p_1^{(i)}\end{equation} 
is the probability of generating ideal evolution throughout the entire circuit. It can be shown that $F_0$ is a lower bound to the quantum state fidelity, from which we can derive an upper bound $M = \log(1-P)/\log(1-F_0)$ on how many measurements $M$ are necessary to sample a result from the ideal quantum probability distribution with probability $P$ \cite{lotshaw2022scaling}.  For example, if the ideal circuit prepared the optimal solution, then in the absence of noise a single measurement would provide this optimal solution.  In the presence of noise, if we wanted to sample this solution with probability $P$, then we would need at most $M$ measurements.

We are now ready to consider how sparsification is expected to influence noise in digital quantum circuits.  As a first approximation, we can consider that all two-qubit gates $U_{uv}$ have identical probabilities for ideal evolution $\overline p_1$, and we can neglect single-qubit gate errors, since these are typically much smaller than two-qubit gate errors. A more precise treatment can take $\overline p_1$ as the geometric mean of two-qubit gate errors. In either case we have $F_0 = \overline p_1^{2m}$ for a graph with $m$ edges, since each edge requires two two-qubit gates in our compilation. For dense instances $m$ may scale quadratically with $n$ while for sparsified instances with a constant error tolerance $\epsilon$ the number of sparsified edges $m' = O(n \log n)$ scales nearly linearly with $n$. The fidelity lower bounds for these circuits satisfy $F_0' = F_0^{m'/m}$, which may be a very significant improvement in cases where $m' \ll m$.  Minimizing the number of noisy operations through sparsification is therefore expected to increase the circuit fidelity and to reduce the number of measurements needed for a high quality result.

\subsection{Sparsified and decomposed compilations of QAOA for analog quantum computation}

Here we consider sparsification and decomposition in compilations for analog trapped-ion quantum hardware, with a specific noise model, as follows.  We consider an $n$-ion quantum state evolving continuously in time under a native optical-dipole-force interaction described by an effective Ising Hamiltonian $H = n^{-1}H_\text{Ising}$ \cite{bohnet2016quantum}; the scaling $\sim n^{-1}$ is a realistic feature that arises from coupling to the center-of-mass vibrational mode. The $\sigma^x_u$ operations are implemented through $\pi$ pulses, which we approximate as instantaneous and noiseless. For the evolution under $H$ we model noise as dephasing on each qubit at rate $\Gamma$. Such dephasing is present in analog trapped ion experiments, due to fundamental Rayleigh scattering of photons as well as technical noise sources, which contributes significantly to single-qubit decoherence in experiments such as Refs.~\cite{bohnet2016quantum,uys2010decoherence}.  This is one of several sources of noise that are present in large-scale trapped ion experiments, and we choose this particular noise model because it is amenable to an exact analytic treatment of single-layer QAOA compiled with \unionofstars, with or without sparsification and decomposition, as described further in Appendix \ref{derivation appendix}.  

We emphasize that our dephasing noise model does not capture all sources of noise that are relevant in trapped ion experiments. Additional sources of noise, such as Raman light scattering transitions, laser power fluctuations, electron-vibration coupling, and state preparation and measurement errors are also important in first-principles physical models of trapped ion dynamics \cite{foss2013nonequilibrium,lotshaw_modeling_2023,lotshaw2024exactly,monroe2021programmable}. For example, M{\o}lmer-S{\o}rensen interactions are often timed or controlled so that the periodic vibrational modes return to their initial states and become disentangled from the electrons after the interaction \cite{blumel2021power,valahu2022quantum}. However the complexity of the time-dependent vibrational spectrum presents challenges for gate timing and control, which can lead to undesired entanglement between the electronic and vibrational degrees of freedom with a decreased purity of the computational state \cite{lotshaw_modeling_2023}.  This source of noise could be expected to decrease computational fidelity in a way that depends on the number of applications of $H$ in the graph compilation, since each application of $H$ provides an opportunity for generating vibration-electron entanglement.  
Methods to address these and other error sources are topics of ongoing research \cite{monroe2021programmable,valahu2022quantum}. However, treating these additional sources of noise analytically in the present context requires a considerably more complicated analysis that is beyond our scope here. We instead focus on dephasing noise only, which we are able to treat analytically in single-layer QAOA.  Although this is only a simplified treatment, it nonetheless enables us to understand some experimentally relevant effects of noise for large instances and deep compilations, which would not be possible with detailed physics simulations of all relevant noise sources.

We model the continuous-time quantum evolution using the Lindbladian master equation 
\begin{equation} \frac{d\rho}{dt} = -i (H\rho - \rho H) - \sum_u (J_u J_u^\dag \rho - 2J_u \rho J_u^\dag + \rho J_u J_u^\dag ). \end{equation}
Here $-i(H\rho - \rho H)$ represents the ideal (Schr\"odinger) quantum state evolution while the $J_u = \sqrt{\Gamma/8}\sigma^z_u$ represent noise due to dephasing, and $H = n^{-1}H_\mathrm{Ising} = n^{-1}\sum_{i < j} \sigma^z_i \sigma^z_j$ is the effective Hamiltonian for the optical-dipole-force detuned close to the $n$-ion center-of-mass mode \cite{bohnet2016quantum,foss2013nonequilibrium}. Using the techniques from Refs.~\cite{foss2013nonequilibrium,lotshaw2024exactly}, we derived the cost expectation value for single-layer QAOA with problem graph coupling operator $C = \sum_{(u,v) \in E} c_{uv} \sigma_u^z \sigma_v^z$ and with a potentially different graph coupling operator $C' = \sum_{(u,v) \in E'} c_{uv}' \sigma_u^z \sigma^z_v$ in compilation, which can represent the approximate coupling operators used in sparsification or decomposition.  The cost expectation value is
\begin{widetext}
\begin{align} \label{<C>} \langle C\rangle & = \sum_{u<v} \frac{c_{uv}\sin(4\beta)\sin(2\gamma(t) c_{uv}')e^{-\Gamma t/2}}{2} \left( \prod_{\mu\neq u,v}\cos(2\gamma(t) c_{\mu v}') + \prod_{\mu\neq u,v}\cos(2\gamma(t) c_{\mu u}') \right) \nonumber \\
& - \sum_{u<v} \frac{c_{uv}\sin^2(2\beta)e^{-\Gamma t}}{2} \left(\prod_{\mu \neq u,v} \cos(2\gamma(t)(c_{\mu u}' + c_{\mu v}')) - \prod_{\mu \neq u,v} \cos(2\gamma(t)(c_{\mu u}' - c_{\mu v}'))\right), \end{align}
\end{widetext}
where $t = \gamma T$ is the total amount of time that the system evolves under the $H_\mathrm{Ising}$ pulses during execution of the algorithm, $\gamma$ and $\beta$ are variational parameters of the algorithm, and $T$ (eqn.~(\ref{eqn: time-of-compilation})) is the amount of time it takes to compile the unitary $\exp(-i C')$.  In the noiseless limit $\Gamma \to 0$, the expression (\ref{<C>}) agrees with the generic QAOA expectation value in eqn.~(14) of Ref.~\cite{ozaeta2022expectation}, while the value is exponentially suppressed when $\Gamma > 0$. We focus here on single-layer QAOA because we are able to evaluate its performance at large sizes using the analytical formula (\ref{<C>}). Greater numbers of QAOA layers would be expected to yield higher performance in the noiseless limit, but closed-form expressions for generic QAOA instances at greater numbers of layers have not been presented in the literature to our knowledge, and we have not derived such formulas in the noisy cases analyzed here. Noiseless analysis of QAOA at larger depths is possible in certain highly symmetric instances \cite{basso2022quantum,farhi_quantum_2022}, but extending our noisy analysis to these cases is beyond the scope of this work.

\begin{figure*}\includegraphics[height=10cm,width=\textwidth,keepaspectratio]{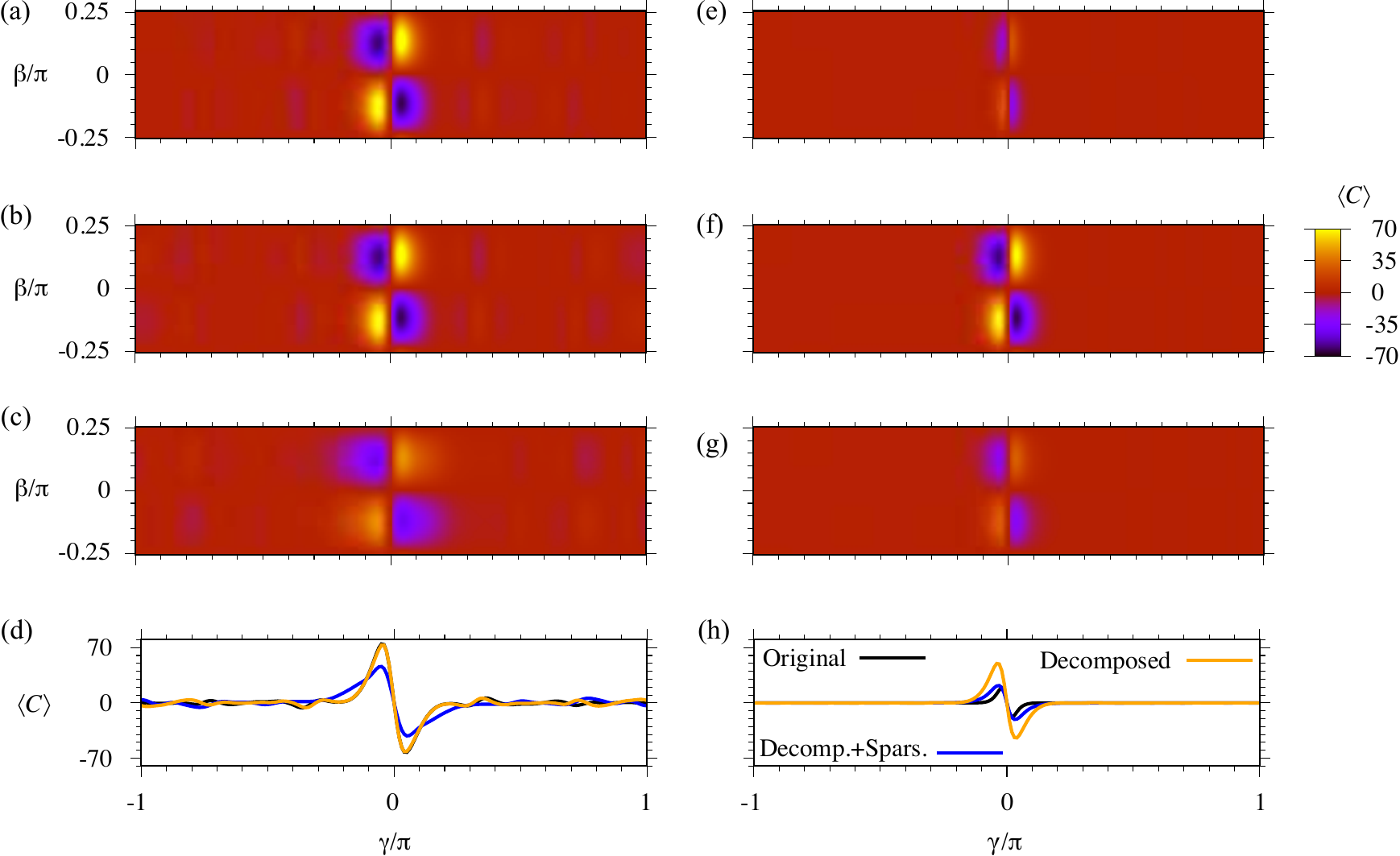}
    \caption{Landscapes of the QAOA cost function.  Left column shows the noiseless case $\Gamma=0$ for (a) standard compilation, (b) compilation with decomposition, (c) compilation with decomposition and sparsification, and (d) line traces through the optimal $\beta$ for each contour in (a)-(c). The right column (e)-(h) shows analogous results in a noisy case with $\Gamma=0.001$, and have the same vertical axes with the figures in the left column. The instance shows the median improvement to $\langle C\rangle_\text{sparsified+decomposed}/\langle C\rangle_\text{original} = 1.1$, and since the optimal $\langle C\rangle_\text{original} < 0$, this indicates that $\langle C\rangle_\text{sparsified+decomposed} <\langle C\rangle_\text{original}$ and therefore the sparsified and decomposed compilation outperforms the original one in minimizing the expected cost (recall eqn. (\ref{eqn: operator-expectation-vs-max-cut-values}) and subsequent discussion); the compilation with decomposition alone achieves an even greater benefit. The graph identifier for this instance is \texttt{g001098} with sparsification parameter $q = 0.5m$ and $\epsilon_2 = 2$, where $m$ is the number of edges in the graph.}
    \label{landscapes}
\end{figure*}

Fig. \ref{landscapes} shows an example of the cost landscape $\langle C\rangle$ as a function of the QAOA parameters $\gamma$ and $\beta$, for cost functions that have been normalized to set the average edge weight to unity following Ref.~\cite{shaydulin2023parameter}.  The left column shows noiseless cases of (a) the original compilation \cite{rajakumar_generating_2022}, (b) decomposed compilation, and (c) decomposed and sparsified compilation, with (d) showing a line cut through the optimal $\beta$.  The results with decomposition are essentially identical to the original compilation, while a much larger error is incurred by the sparsified compilation because the approximate cuts in $C'$ produce an approximate quantum state from the circuit.   (e), (f), (g), and (h) show analogous results for the same instance in the presence of noise, with $\Gamma=0.001$.  Here the decomposed compilation greatly outperforms the original compilation, while the sparsified and decomposed compilation slightly outperforms it. 
 
One surprising feature of these results is the extent to which sparsification is harming the performance.  In the noiseless case, the lower approximation ratio in Fig.~\ref{landscapes} is due to the approximate graph coupling $C'$ as mentioned previously.  In the noisy cases, there is still little or no benefit in our calculations because sparsification is not correlated with reduced execution time $t$ as seen previously in Fig.~\ref{fig: time-of-pulses-sparsification}, while noise in our model depends only on the execution time in eqn.~(\ref{<C>}).  However, it is important to keep in mind that a more realistic noise model may see benefits from sparsification which are not evident in the simple model we consider here.  For example, if there is some residual entanglement between the vibrational and electronic modes after each application of $H$, as described above and as expected in more realistic treatments of trapped-ion dynamics \cite{lotshaw_modeling_2023}, then we would expect that limiting the number of $H_\text{Ising}$ applications through sparsification (Fig.~\ref{fig: number-of-pulses-sparsification}) would produce a less noisy final result.  Similar considerations also apply if there is noise associated with bit flip operations (Fig.~\ref{fig: total-gates-sparsification}).  This gives some reason to expect that sparsification may still be useful for limiting noise in real-world trapped ion experiments.  Finally, it is important to note that the analog trapped-ion noise model we consider in this section is distinct from the digital quantum-computing considerations from the previous section, where there is a direct link between sparsification, the number of edges, and the expected reduction in noise. 

\begin{figure*}
    \centering
    \includegraphics[height=7cm,width=\textwidth,keepaspectratio]{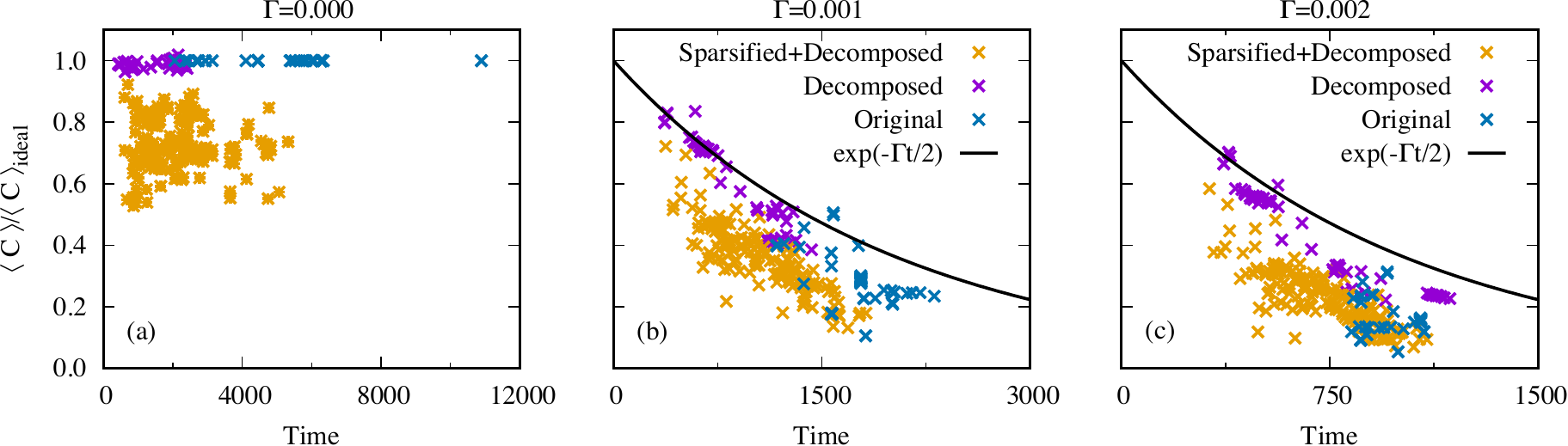}
    \caption{Expected cost $\langle C \rangle$ in QAOA for original, decomposed, and decomposed and sparsified and sparsified compilations, relative to the "ideal" case without noise and with the original compilation. (a) noiseless case, (b) noisy case with $\Gamma=0.001$, and (c) with $\Gamma=0.002$.  The black curve $\exp(-\Gamma t/2)$ is an upper bound in the triangle-free case as described in the text. Each instance is compiled into three distinct sparsified compilations. All panels share a common vertical axis.  }
    \label{fig:cost}
\end{figure*}

Next, we examine the optimized performance across a wide variety of instances using the original compilation, decomposition, and for three implementations of sparsification + decomposition for each instance. The standard procedure for implementing QAOA includes optimizing the parameters $\bm \gamma$, $\bm \beta$ for high performance \cite{farhi_quantum_2014}. Many methods for accomplishing this have been studied \cite{shaydulin2023parameter,lotshaw2021empirical,lotshaw2023approximate,zhou_quantum_2020,farhi_quantum_2022,augustino2024strategies,wurtz2021fixed}. Among these grid searches \cite{farhi_quantum_2014,lotshaw2023simulations} are simple options that are guaranteed to find optimal parameters within the relevant parameter subspace and to within the resolution of the grid spacing. We use grid searches here to obtain these optimal parameters, so there is no uncertainty in the results due to the parameter optimization procedure, allowing us to directly assess the best possible performance, within a reasonable parameter interval, for the various decomposition methods. We identify optimized QAOA parameters using a grid search with resolution of $0.01\pi$ in $\gamma$ and $\beta$, over the parameter regions shown in Fig.~\ref{landscapes} where high-quality results are expected \cite{shaydulin2023parameter}. 

In Fig.~\ref{fig:cost} we show the optimized performance in (a) the noiseless case, (b) a noisy case with $\Gamma=0.001$, and (c) with $\Gamma=0.002$. The decomposed compilation is comparable to the original compilation in the noiseless case and significantly outperforms it in the noisy case, while sparsification together with decomposition yields more modest improvements, all as expected from our previous analysis of Fig.~\ref{landscapes}. To gain a better understanding of the results in Fig.~\ref{fig:cost}, next we will analyze the expected noisy cost under a simplifying assumption, which will explain the approximate upper bound $\exp(-\Gamma t/2)$ pictured in black. 
 
The expected cost eqn.~(\ref{<C>}) can be expressed as $\langle C \rangle = e^{-\Gamma t/2} f(\gamma,\beta) + e^{-\Gamma t} \tau(\gamma,\beta)$, where the function $f$ [given by the first set of sums in eqn.~(\ref{<C>})] describes contributions from each edge $(u,v)$ in the graph while the function $\tau$ [given by the second set of sums in eqn.~(\ref{<C>})] describes additional contributions when there are triangles in the graph.  We find that for each of our compilation approaches, the optimized contribution of the triangle term is typically $|\tau(\gamma^*,\beta^*)| <  |f(\gamma^*,\beta^*)|/10$.  We can therefore approximate $\langle C \rangle \approx e^{-\Gamma t/2}f(\gamma^*,\beta^*)$.  We would now like to consider noisy behavior relative to noiseless behavior, starting with the original compilation only.  We then have $\langle C \rangle_\text{noise}/\langle C\rangle_\text{ideal} \approx e^{-\Gamma t^{'*}/2} f(\gamma^{'*},\beta^{'*})/f(\gamma^{*},\beta^{*})$, where $\gamma^{'*},\beta^{'*}$ are the chosen parameters in the presence of noise.  In the simple fixed-parameter case $(\gamma^{'*},\beta^{'*}) = (\gamma^{*},\beta^{*})$ this reduces to $\langle C \rangle_\text{noise}/\langle C\rangle_\text{no\ noise} = e^{-\Gamma t^{'*}/2}$.  However, as we saw previously in Fig.~\ref{landscapes}, noise tends to suppress the optimal parameter such that $\gamma^{'*} < \gamma^*$.  In this case $\langle C \rangle_\text{noise}/\langle C\rangle_\text{ideal} = e^{-\Gamma t^{'*}/2} f(\gamma^{'*},\beta^{'*})/f(\gamma^{*},\beta^{*}) < e^{-\Gamma t^{'*}/2}$, since $ f(\gamma^{'*},\beta^{'*})/f(\gamma^{*},\beta^{*}) < 1$ as $\gamma^{'*}$ is not the ideal parameter $\gamma^*$ that maximizes $f$.  Note that $t^{'*} \propto\gamma^{'*}$ and is smaller in this second case than in the $\gamma^{'*}=\gamma^*$ case.  Based on this simple model, we expect that with the original compilation, $\langle C \rangle_\text{noise}/\langle C\rangle_\text{ideal} \lesssim e^{-\Gamma t/2}$ for an optimized noisy runtime $t$, with a strict inequality $\langle C \rangle_\text{noise}/\langle C\rangle_\text{ideal} \leq e^{-\Gamma t/2}$ in the triangle-free case.  When we use other compilations, then $\langle C \rangle_\text{noise}$ may decrease further due to the use of approximate compilation.  So in all cases we expect $\langle C \rangle_\text{noise}/\langle C\rangle_\text{ideal} \lesssim e^{-\Gamma t/2}$.  This simple model gives a good account of the typical behavior observed in Fig.~\ref{fig:cost}.
    
\begin{figure*}
    \centering
    \includegraphics[height=9cm,width=9cm,keepaspectratio]{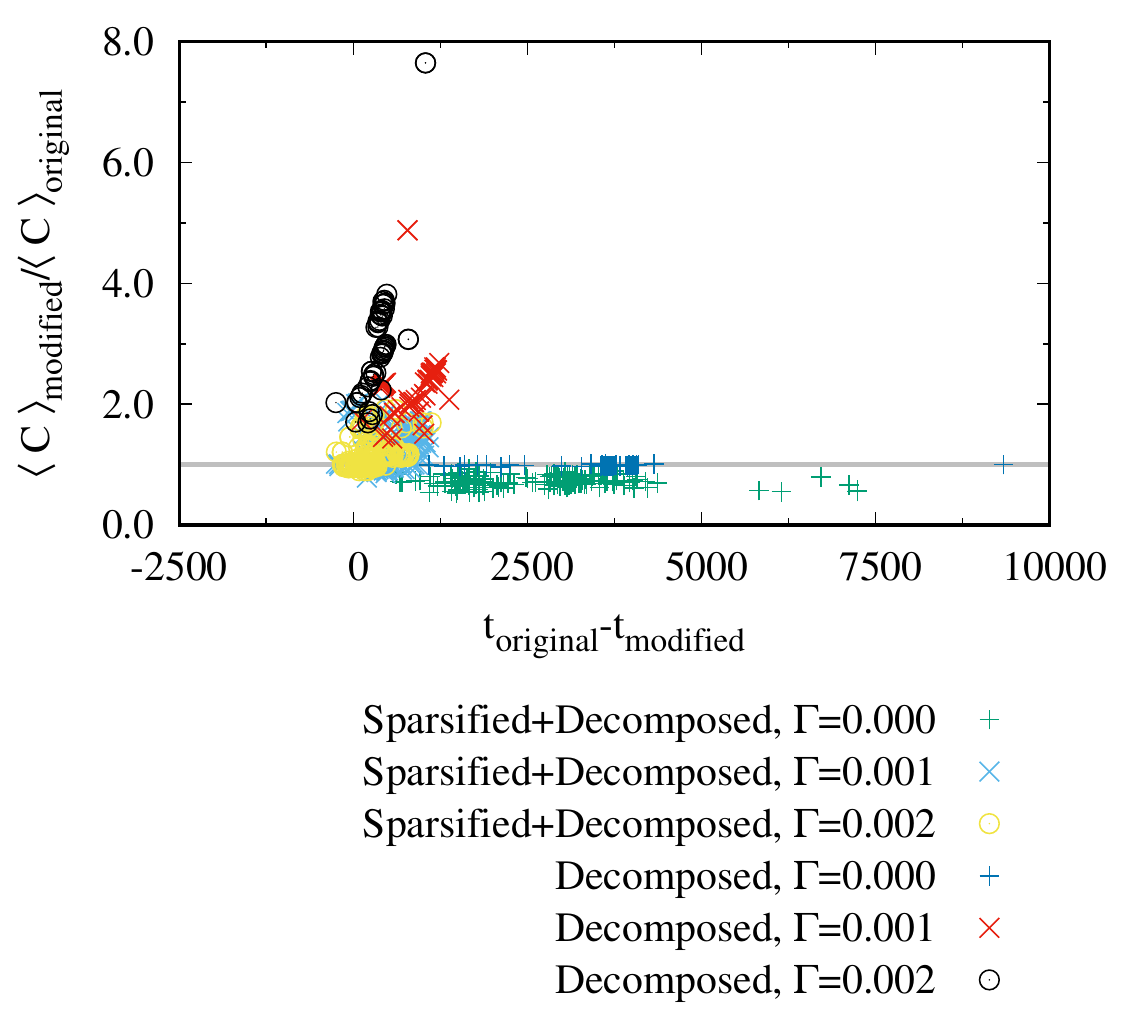}
    \caption{Expected improvement in cost from modified compilations relative to the original compilation.  The grey line shows $\langle C\rangle_\text{modifed}/\langle C\rangle_\text{original}=1.$  }
    \label{fig:costratio}
\end{figure*}

Finally, in Fig.~\ref{fig:costratio} we give a direct comparison of the costs from our various compilations relative to the original compilation.  In the presence of noise, our new compilations consistently outperform the original compilation.  This result provides strong evidence that our new compilation approaches here provide superior performance to the original compilation when noise is considered.

\section{Conclusion}\label{sec: conclusion}

Noise in quantum hardware presents a significant hurdle to achieving scalable quantum computing. We presented a problem-aware approach for noise reduction in QAOA that modifies the problem instance suitably, resulting in shallower circuit depth and therefore less noise.
We presented theoretical bounds and numerical evidence for reduction in the number of circuit gates while guaranteeing high Max-Cut approximations on trapped-ion hardware employing all-to-all M{\o}lmer-S{\o}rensen interactions. We also presented a generic theoretical argument for how these techniques can improve the fidelity of digital quantum computations with superconducting qubits \cite{bravyi2022future}, Rydberg atoms \cite{bluvstein2024logical}, trapped-ions employing two-qubit gates \cite{pino2021demonstration}, or any other digital quantum computing technology with nonnegligible noise described by Krauss operators.
This approach reduces circuit noise and allows QAOA to scale for larger graphs on NISQ hardware, thus narrowing the performance gap between classical and quantum hardware. However, quantum noise remains a major challenge, and classical methods continue to outperform quantum algorithms for combinatorial optimization.

Other approaches to noise reduction in QAOA have been developed in parallel to our work. These include finding smaller graphs with similar energy landscapes \cite{wang_red-qaoa_2024}, integer programming-based approaches for crosstalk noise reduction \cite{wang_red-qaoa_2024}, and using machine learning approaches to predict QAOA parameters \cite{sack_large-scale_2024}. To the best of our knowledge, ours is the only approach with formal worst-case guarantees on the approximation ratio.

While most of our asymptotic theoretical guarantees are restricted to trapped-ion hardware, both sparsification and decomposition are general tools that may be useful beyond the trapped-ion setting.
Sparsification is expected to improve noise resilience in generic compilations since it simplifies the problem instance, resulting in fewer gates and an exponentially improved fidelity lower bound as described here.
Further, since there exist faster classical combinatorial algorithms with better guarantees for unweighted instances, relative to weighted instances \cite{cormen2022introduction, williamson_design_2010}, we believe decomposition-like techniques might be useful in analyzing quantum optimization algorithms for weighted instances.

\section{Author contributions}

This work sits at the intersection of classical and quantum computing: a central challenge was to identify quantum noise models that admit rigorous analysis alongside classical graph sparsification and decomposition techniques. All authors contributed to the design of the study and experiments. Swati Gupta (SG) and Jai Moondra (JM) introduced the idea of classical graph pre-processing to generate QAOA compilations with smaller circuits. Phillip C. Lotshaw (PCL) and Greg Mohler (GM) introduced the quantum noise models, with PCL leading the noise analysis and the derivation of the dephased QAOA cost expectation. JM led the analysis of the sparsification and decomposition algorithms and the resulting reductions in gate counts; the theoretical guarantees were developed jointly by JM and SG with input from all authors. PCL ran all quantum simulations; JM performed the classical sparsification and decomposition experiments, both with input from all authors. All authors contributed to writing and editing the manuscript.

\paragraph{Acknowledgements.} This material is based upon work supported by the Defense Advanced Research Projects Agency (DARPA) under Contract No. HR001120C0046.
The authors would like to thank Brandon Augustino, Cameron Dahan, Bryan Gard, Mehrdad Ghadiri, Creston Herold, Sarah Powers, and Mohit Singh for their careful comments and helpful discussions on this work.

\normalsize
\theendnotes

\bibliographystyle{plain}

\newpage

\appendix

\section{Omitted proofs from Section \ref{sec: weighted-graphs-faster-union-of-stars}}\label{sec: algorithms-analysis}

We analyze our algorithms for faster \unionofstars\ from Section \ref{sec: weighted-graphs-faster-union-of-stars} and prove their theoretical guarantees.

\subsection{Analysis of Algorithm \customnameref{alg: exponential-decomposition}}\label{app:exp-decompose}

\begin{proof}[Proof of Theorem \ref{thm: exp-decompose}]
    Recall that in Algorithm \customnameref{alg: exponential-decomposition}, we set $c^* = \max_{e \in E}c(e)$ and threshold $\tau = \frac{\epsilon c^*}{2 n^2}$. Correspondingly we set $k = \lceil \log_{1 + \epsilon/2} \tau \rceil = O\left(\frac{\log(n/\epsilon)}{\epsilon}\right)$.
    
    Consider a cut $S \subseteq V$. By construction in the algorithm, $c'(e) \le c(e)$ for all edges $e \in E$, and therefore the cut value can only decrease in $G' = (V, E', c')$, i.e., $\delta_{G'}(S) \le \delta_G(S)$. We next show that it does not decrease by too much, i.e., $\delta_{G'}(S) \ge (1 - \epsilon) \delta_G(S)$ if $S$ is a non-trivial cut in $G$, i.e., if $\delta_G(S) \ge \frac{1}{2} \sum_{e \in E} c(e)$.

    Fix edge $e \in E$. Denote $E_{\text{small}} := \{e \in E: c(e) < \tau\}$ and $E_{\text{large}} := \{e \in E: c(e) \ge \tau\}$. If $e \in E_{\text{small}}$, then $e' \not \in E'$ by construction in the algorithm. If $e \in E_{\text{large}}$, then by construction in line \ref{step: exp-decompose-edge-weight-multiplicative-approximation}, we get $c'(e) \ge \left(1 - \frac{\epsilon}{2}\right) c(e)$.
    
    Noting that $\Delta_{G'}(S) = \Delta_{G}(S) \cap E_{\text{large}}$, the decrease in the cut value for $S$ can be bounded as:
    \begin{widetext}
    \begin{align*}
        \delta_G(S) - \delta_{G'}(S) &= \sum_{e \in \Delta_{G}(S)} c(e) -  \sum_{e \in \Delta_{G'}(S)} c'(e) \\
        &= \sum_{e \in \Delta_{G}(S) \cap E_{\text{small}}} c(e) + \sum_{e \in \Delta_{G}(S) \cap E_{\text{large}}} c(e) - \sum_{e \in \Delta_{G}(S) \cap E_{\text{large}}} c'(e) \\
        &\le \tau |\Delta_G(S) \cap E_{\text{small}}| + \frac{\epsilon}{2} \sum_{e \in \Delta_{G}(S) \cap E_{\text{large}}} c(e).
    \end{align*}
    \end{widetext}
    We will claim that each of the terms above is at most $\frac{\epsilon}{2} \delta_G(S)$, which is sufficient to prove part (1) of the theorem.

    It is easy to see that $\frac{\epsilon}{2} \sum_{e \in \Delta_{G}(S) \cap E_{\text{large}}} c(e) \le \frac{\epsilon}{2} \sum_{e \in \Delta_{G}(S) } c(e) = \frac{\epsilon}{2} \delta_G(S)$. For the first term, plugging in $\tau = \frac{\epsilon c^*}{2 n^2}$,
    \begin{align*}
        & \ \tau |\Delta_G(S) \cap E_{\text{small}}| = \frac{\epsilon c^*}{2 n^2} |\Delta_G(S) \cap E_{\text{small}}| \\
        \le \ & \frac{\epsilon c^*}{2 n^2} |E| \le \frac{\epsilon c^*}{2 n^2} \cdot \frac{n^2}{2} \le \frac{\epsilon}{4} c^*.
    \end{align*}
    However, $c^* \le \sum_{e \in E} c(e) \le 2 \delta_G(S)$, since $S$ is a non-trivial cut by assumption. Therefore, $\frac{\epsilon}{4} c^* \le \frac{\epsilon}{2} \delta_G(S)$. This completes the proof of the claim.

    To see part (2), note that by construction, $E'$ is the disjoint union of $\bigcup_{j \in [k]} E_j$. Further, each of $G = (V, E_j)$ is an unweighted graph. By Lemma \ref{lem: union-of-stars-graph-combination} and Lemma \ref{thm: undirected-graph-gc}, we have that 
    \begin{align*}
        & \nHising(G') \le k \cdot (3n - 2) \\
        = \ & \left(\log_{1 + \frac{\epsilon}{2}}\frac{2n^2}{\epsilon} \right) \cdot (3n - 2) = O\left(\frac{n}{\epsilon} \log \frac{n}{\epsilon}\right). \qedhere
    \end{align*}
\end{proof}

\subsection{Analysis of Algorithm \customnameref{alg: binary-decomposition}}

\begin{proof}[Proof of Theorem \ref{thm: faster-union-of-stars-weighted-graphs}.]
    Denote $G = (V, E, c)$. Let $c^*, \eta, k$ be as in Algorithm \customnameref{alg: binary-decomposition}. Recall that for each $e \in E$, we define $d(e) = \lfloor \frac{c(e)}{\eta} \rfloor$ for each $e \in E$, and $b_{k - 1}(e) b_{k - 2}(e) \ldots b_0(e)$ are the digits in the binary representation of $d(e)$. Therefore, $d(e) = \sum_{j \in [k]} 2^{j - 1} b_{j - 1}(e)$.
    
    Let $G_1, \ldots, G_k$ be the unweighted graphs in Algorithm \customnameref{alg: binary-decomposition}. Recall that the algorithm returns $G' = (V, E', c')$ such that $G' = \sum_{j \in [k]} \eta 2^{j - 1} G_j$.

    \begin{enumerate}[label=(\roman*)]
        \item For each edge $e \in E$, the edge weight in $G'$ is $c'(e) = \sum_{j \in [k]} \eta 2^{j - 1} b_{j - 1}(e) = \eta d(e)$. Then we get
        \[
            c'(e) = \eta d(e) = \eta \left\lfloor \frac{c(e)}{\eta} \right\rfloor \ge \eta \left(\frac{c(e)}{\eta} - 1\right) = c(e) - \eta.
        \]
        Also, $c'(e) \le c(e)$. Let $S \subseteq V$ be any set of vertices; there are at most $n^2/2$ edges in the corresponding cut. Summing across all edges in the cut, we get
        \[
            \delta_{G}(S) - \delta_{G'}(S) \le |E| \eta = |E|\frac{c^*\epsilon}{n^2} \le \frac{c^*}{2} \epsilon \le \left(\frac{1}{2} \sum_{e \in E} c(e)\right) \epsilon.
        \]
        Since $S$ is a non-trivial cut, $\delta_G(S) \ge \frac{1}{2} \sum_{e \in E} c(e)$, so that $\delta_{G}(S) - \delta_{G'}(S) \le \epsilon \delta_G(S)$. That is, $\delta_{G'}(S) \ge (1 - \epsilon)\delta_G(S)$.

        In particular, if $\hat{S}, \hat{S}'$ are the vertex sets corresponding to Max-Cut in $G, G'$ respectively, we have
        \[
            \delta_G(\hat{S}') \ge \delta_{G'}(\hat{S}') \ge \delta_{G'}(\hat{S}) \ge (1 - \epsilon) \delta_G(\hat{S}).
        \]
        The first inequality holds since $c'(e) \le c(e)$ for all $e \in E$, the second inequality holds since $\hat{S}$ is the Max-Cut in $G'$, and the third inequality holds since $\hat{S}$ is a non-trivial cut. Therefore, the Max-Cut $\hat{S}'$ in $G'$ is a $(1 - \epsilon)$-approximate cut in $G$.
        
        \item Since each $G_j, j \in [k]$ is an unweighted graph, we have $\nHising(G_j) \le 3n - 2$ from Theorem~\ref{thm: undirected-graph-gc}. Since $G' = \sum_{j \in [k]} \eta 2^{j - 1} G_j$, we get from Lemma~\ref{lem: union-of-stars-graph-combination} that
        \begin{align*}
            & \nHising(G) \le \sum_{j \in [k]} \nHising(G_j) \\
            \le \ & k (3n - 2) = O(n \log(n/\epsilon)).
        \end{align*}
    \end{enumerate}
    Finally, note that each edge in $G'$ is an edge in at least one of $G_1, \ldots, G_k$, and therefore it is an edge in $G$.
\end{proof}

\subsection{Analysis of Algorithm \customnameref{alg: faster-union-of-stars-2}}

\begin{proof}[Proof of Theorem \ref{thm: second-graph-coupling-number}]
    Let $H = (V, \ov{E}, \ov{c})$ be the sparsified graph in line 1 of Algorithm \customnameref{alg: faster-union-of-stars-2}. Then by Theorem \ref{thm: sparsification}, with high probability, for all $S \subseteq V$, $1 - \frac{\epsilon}{3} \le \frac{\delta_H(S)}{\delta_G(S)} \le 1 + \frac{\epsilon}{3}$. Further, $|\ov{E}| = O\left(\frac{n}{\epsilon^2}\right)$.

    \textbf{Case I}. $G' = G'_\text{bin} = \customnameref{alg: binary-decomposition}(H, \epsilon_2)$. From Theorem \ref{thm: faster-union-of-stars-weighted-graphs}, $G'$ satisfies that (1) $\nHising(G') = O\left(n \log (n/\epsilon)\right)$ and (2) $1 - \frac{\epsilon}{3} \le \frac{\delta_{G'}(S)}{\delta_G(S)} \le 1$ for all non-trivial cuts $S \subseteq V$. This latter equation, along with the observation above, gives that for all $\epsilon \in (0, 1]$,
    \begin{align*}
        1 + \frac{\epsilon}{3} \ge \frac{\delta_{G'}(S)}{\delta_G(S)} \ge \left(1 - \frac{\epsilon}{3}\right)^2 = 1 - \frac{2 \epsilon}{3} + \frac{\epsilon^2}{9} \ge 1 - \epsilon.
    \end{align*}

    Let $G_1, \ldots, G_k$ be the graphs in \customnameref{alg: binary-decomposition} for inputs $H, \epsilon/3$. Recall that $k = O(\log(n/\epsilon))$. Since $E(G_j) \subseteq \ov{E}$ for each $j \in [k]$, we get that $\nTotal(G_j) = O(|E(G_j)|) = O\left(\frac{n}{\epsilon^2}\right)$. Therefore, from Lemma \ref{lem: union-of-stars-graph-combination},
    \begin{align*}
        \nTotal(G') &\le \sum_{j \in [k]} \nTotal(G_j) \\
        &= O\left(\frac{n \log(n/\epsilon)}{\epsilon^2}\right).
    \end{align*}

    \textbf{Case II}. $G' = G'_{\text{exp}} = \customnameref{alg: exponential-decomposition}(H, \epsilon_2)$. By the choice of $G'$ and from Theorem \ref{thm: exp-decompose}, $G'$ satisfies that $\nHising(G') = \nHising(G'_{\text{exp}}) \le \nHising(G'_\text{bin}) = O\left(n \log (n/\epsilon)\right)$.  
    
    Let $G_1, \ldots, G_k$ be the graphs in \customnameref{alg: exponential-decomposition} for inputs $H, \epsilon/3$. Recall that $k = O\left(\frac{1}{\epsilon}\log(\frac{n}{\epsilon})\right)$. Since $E(G') \subseteq E(H)$ is the disjoint union of $E(G_j), j \in [k]$, we get that 
    \begin{align*}
        & \sum_j \nTotal(G_j) = \sum_j O(|E(G_j)|) \\
        = \ & O(|E(H)|) = O\left(\frac{n}{\epsilon^2}\right).
    \end{align*}
    Therefore, from Lemma \ref{lem: union-of-stars-graph-combination},
    \[
        \nTotal(G') \le \sum_{j \in [k]} \nTotal(G_j) = O\left(\frac{n}{\epsilon^2}\right). 
    \]
    Finally, from Theorem \ref{thm: exp-decompose}, $1 - \frac{\epsilon}{3} \le \frac{\delta_{G'}(S)}{\delta_{G}(S)} \le 1$ for all non-trivial cuts $S$. Therefore, for all such cuts, as before,
    \begin{align*}
        1 + \frac{\epsilon}{3} \ge \frac{\delta_{G'}(S)}{\delta_G(S)} \ge \left(1 - \frac{\epsilon}{3}\right)^2 = 1 - \frac{2 \epsilon}{3} + \frac{\epsilon^2}{9} \ge 1 - \epsilon. \quad \qedhere
    \end{align*}
\end{proof}

\section{A comparison of the two decomposition algorithms}\label{app: bin-decompose}

We presented two algorithms for decomposition in Section \ref{sec: faster-gc-1}: \customnameref{alg: binary-decomposition} and \customnameref{alg: exponential-decomposition}. Given decomposition error parameter $\epsilon_2$, both algorithms improve $\nHising$ for weighted graphs from $O(m) = O(n^2)$ to $O\left(\frac{n}{\epsilon_2} \log \frac{n}{\epsilon_2}\right)$ (for \customnameref{alg: exponential-decomposition}) or to $O\left(n \log \frac{n}{\epsilon_2}\right)$ (for \customnameref{alg: binary-decomposition}). While the two are similar in terms of dependence on $n$, the latter depends only logarithmically on $1/\epsilon_2$ while the former depends linearly on $1/\epsilon_2$. Therefore, one would expect that \customnameref{alg: binary-decomposition} would outperform \customnameref{alg: exponential-decomposition}.

\begin{figure*}[t]
    \begin{minipage}[t]{0.49\textwidth}
        \centering
        \includegraphics[width=\textwidth]{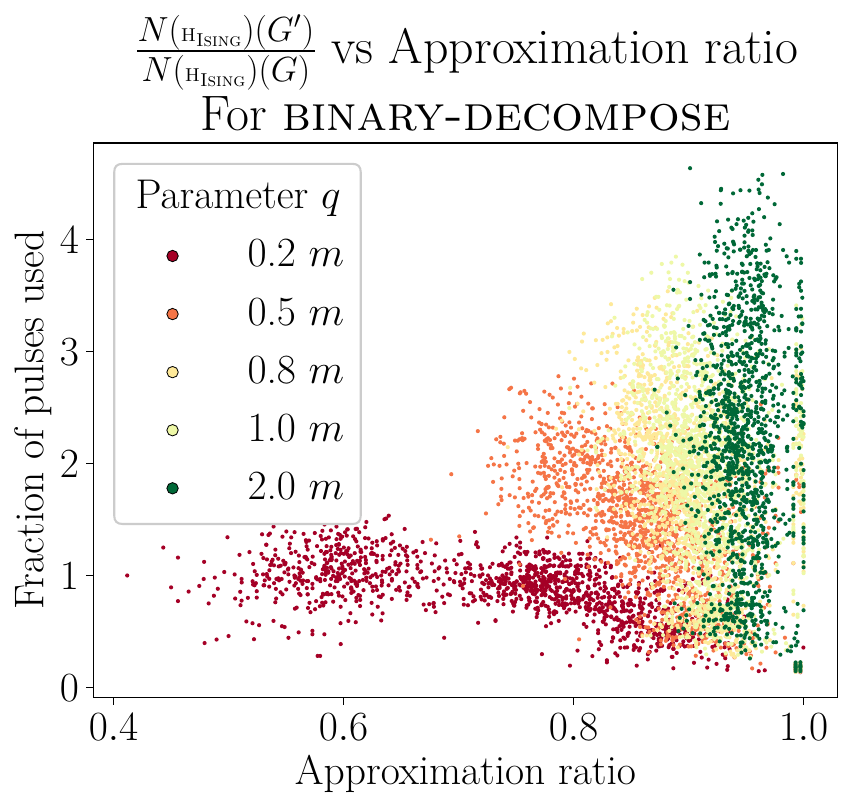}
        \caption{$\nHising(G')/\nHising(G)$ (total number of \hising pulses) vs Max-Cut approximation for various runs of our experiment on weighted graphs in MQLib using \customnameref{alg: binary-decomposition} as the decomposition algorithm, colored by parameter $q$. Each data point is a single run.}
        \label{fig: bin-decompose-gc-1}
    \end{minipage}
    \hfill
    \begin{minipage}[t]{0.49\textwidth}
        \centering
        \includegraphics[width=\textwidth]{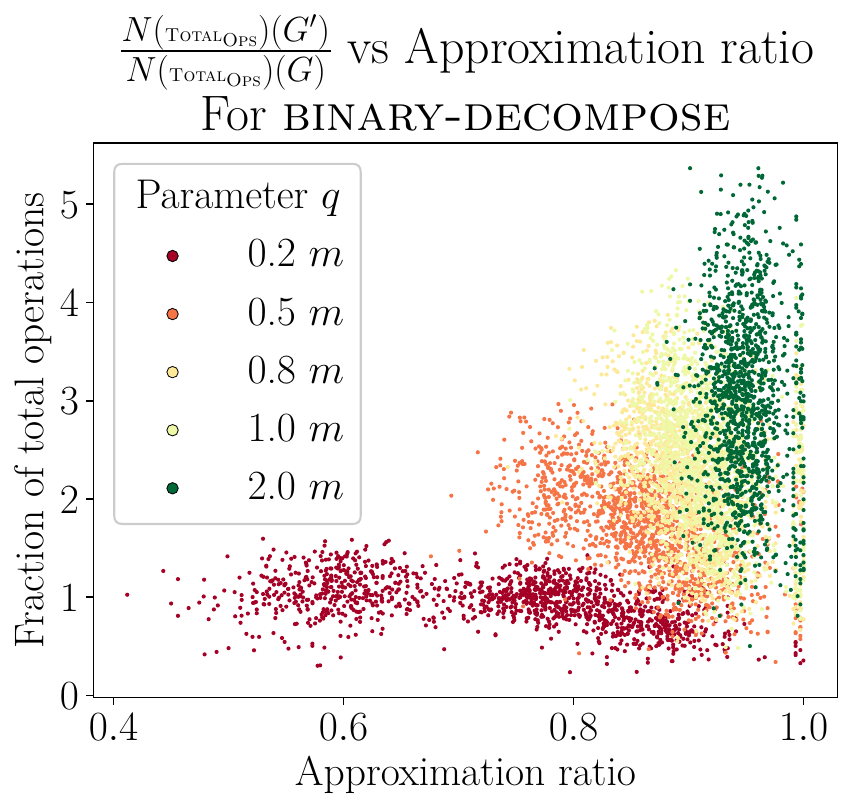}
        \caption{$\nTotal(G')/\nTotal(G)$ (total number of \hising pulses and bit flip gates) vs Max-Cut approximation for various runs of our experiment on weighted graphs in MQLib using \customnameref{alg: binary-decomposition} as the decomposition algorithm. Each data point is a single run.}
        \label{fig: bin-decompose-gc-2}
    \end{minipage}
\end{figure*}

As we noted in our experiments on MQLib graphs, this is usually \emph{not} the case for smaller graphs and for relatively large $\epsilon_2$. This is because of the large constant for \customnameref{alg: binary-decomposition} hidden in the $O$ notation. Indeed, for medium-sized graphs such as the ones we use for our experiments in Section \ref{sec: simulations}, using \customnameref{alg: binary-decomposition} typically \emph{increases} $\nHising$ and $\nTotal$ from base \unionofstars\ (see Figures \ref{fig: bin-decompose-gc-1}, \ref{fig: bin-decompose-gc-2}), as opposed to \customnameref{alg: exponential-decomposition} which significantly decreased these numbers (Figures \ref{fig: number-of-pulses-sparsification} \ref{fig: total-gates-sparsification}).

\begin{figure*}[h]
    \centering
    \includegraphics[width=0.85\textwidth]{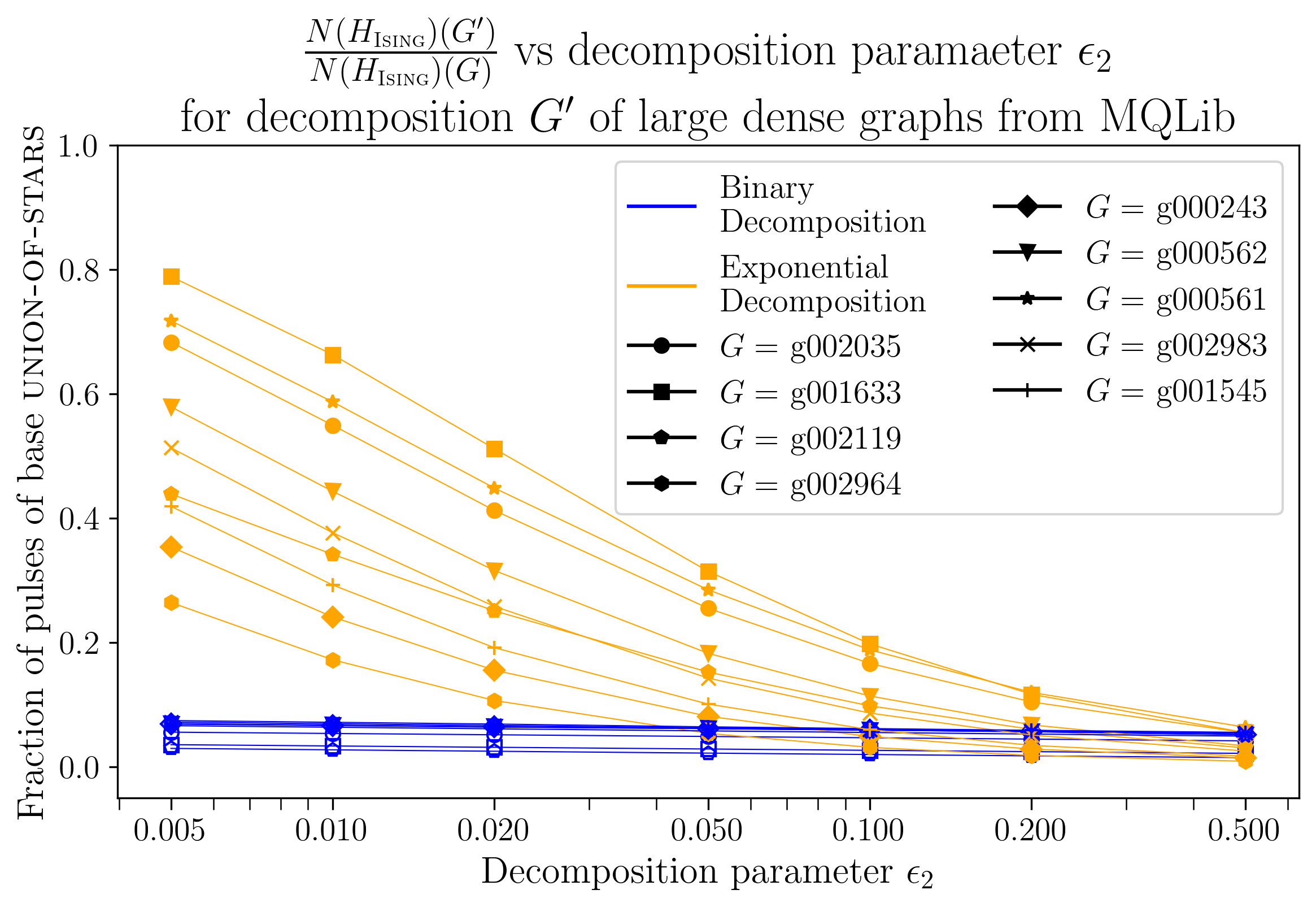}
    \caption{A comparison of $\nHising(G')/\nHising(G)$ (lower is smaller) for decomposed graph $G'$ produced using \customnameref{alg: binary-decomposition} and \customnameref{alg: exponential-decomposition} for nine large and dense graphs in MQLib. Note that as $\epsilon_2 \to 0$, \customnameref{alg: binary-decomposition} starts to outperform \customnameref{alg: exponential-decomposition} significantly. Additionally, for all graphs \customnameref{alg: binary-decomposition} takes fewer than $0.1$ of the \hising pulses taken by base \unionofstars\ even for $\epsilon_2 = 0.005$, i.e., with Max-Cut approximation guaranteed to be preserved by more than $99.5\%$.}
    \label{fig: exp-vs-bin-decompose-large-graphs-others}
\end{figure*}

However, for large and dense enough graphs and for small values of $\epsilon_2$ (i.e., when we seek higher cut quality), \customnameref{alg: binary-decomposition} does outperform \customnameref{alg: exponential-decomposition}. We gave example of one such graph from MQLib in Figure \ref{fig: exp-vs-bin-decompose-large-graph}; Figure \ref{fig: exp-vs-bin-decompose-large-graphs-others} presents several more examples. For this plot, we chose the nine densest graphs in MQLib with fewer than $500,000$ edges.

\section{Derivation of the dephased QAOA cost expectation} \label{derivation appendix}

We model noisy compilation using a dephasing master equation that is a special case of the model presented in Foss-Feig {\it et. al} \cite{foss2013nonequilibrium}.  We begin by considering only the native Hamiltonian $H = n^{-1}H_\mathrm{Ising}$, then include the $\sigma_u^x$ from \unionofstars later.  Define a Lindbladian master equation
\begin{equation} \label{Lindbladian} \frac{d\rho}{dt} = -i(H_\mathrm{eff}\rho-\rho H_\mathrm{eff}^\dag) + D(\rho) \end{equation}
with an effective non-Hermitian Hamiltonian
\begin{align} H_\mathrm{eff} = n^{-1}H_\mathrm{Ising}-i\sum_\mathcal{J} \mathcal{J}^\dag \mathcal{J} \end{align}
and dissipator
\begin{equation} \mathcal{D}(\rho) = 2\sum_\mathcal{J} \mathcal{J} \rho \mathcal{J}^\dag \end{equation}
which are each specified in terms of a set of jump operators $\mathcal{J}$.  Specializing to the case of dephasing only, the set of jump operators is defined as 
\begin{equation} \{\mathcal{J}\} = \left\{\sqrt{\Gamma/8} \sigma_u^z\right\} \end{equation} 

We consider the evolution of a master equation using the quantum trajectories approach \cite{foss2013nonequilibrium,plenio1998quantum}.  The evolution is expressed in terms of an ensemble of pure states evolving under the effective Hamiltonian $H_\mathrm{eff}$, with jump operators that are randomly interspersed in the evolution, following a probability distribution that causes a large ensemble of randomly generated pure states to exactly reproduce the time-evolving density matrix from (\ref{Lindbladian}). The probability of applying a jump operator $\mathcal{J}$ at time $t$ is proportional to $\langle \mathcal{J}^\dag \mathcal{J} \rangle$, which is simply a constant for the dephasing noise we consider since ${\sigma_u^z}^\dag \sigma_u^z = 1.$  Following Ref.~\cite{foss2013nonequilibrium}, we will compute spin-spin correlations along a single trajectory, then average these over the ensemble of all trajectories to obtain exact correlation functions from the density matrix. 

A single trajectory $T=\{\mathcal{J}_{u_1}^{(t_1)},\mathcal{J}_{u_2}^{(t_2)},\ldots,\mathcal{J}_{u_K}^{(t_K)}\}$ with jump operators randomly assigned at times $t_1,\ldots,t_K$ will have the form
\begin{align} \ket{\psi(t;T)} & = e^{-i H_{\mathrm{eff}}(t-t_K)}\mathcal{J}_{u_K} e^{-i H_{\mathrm{eff}}(t_K-t_{K-1})}\mathcal{J}_{u_{K-1}} e^{-i H_{\mathrm{eff}}(t_{K - 1}-t_{K-2})}\ldots\mathcal{J}_{u_{1}} e^{-i H_{\mathrm{eff}}(t_1)} \ket{\psi(0)}\nonumber\\
& = \mathcal{T} \left(e^{-i H_{\mathrm{eff}}t} \prod_i \mathcal{J}_{u_i}\right)\ket{\psi(0)}
\end{align}
where the last line defines a time-ordering operator $\mathcal{T}$ to abbreviate the notation.  If we assume there is no noise associated with bit-flip operations $\sigma_u^x$ in a \unionofstars\ compilation, then we can add these into the otherwise continuous evolution under $H_\mathrm{Ising}$ to obtain a \unionofstars\ trajectory 
\begin{equation} \ket{\psi(t;T)} = \mathcal{T} \left(e^{-i H_\mathrm{eff}t} \prod_i \mathcal{J}_{u_i}\left\{ \prod_a\left[ \prod_{u \in B_a} \sigma_u^x\right]\right\} \right)\ket{\psi(0)}\end{equation}
where $B_a$ is the set of qubits which are bit-flipped in the $a$th step of \unionofstars. The key point to simplifying this trajectory is to realize that $\sigma_u^z\sigma_u^x = -\sigma_u^x\sigma_u^z$, so it is equivalent to commute all the jump operators $\mathcal{J}_{u_i}$ to the right to obtain 
\begin{equation} \ket{\psi(t;T)} = \mathcal{T} \left(e^{-i H_\mathrm{eff}t} \left\{ \prod_a\left[ \prod_{u \in B_a} \sigma_u^x\right]\right\} \right)\left(\pm \prod_i \mathcal{J}_{u_i}\ket{\psi(0)}\right)\end{equation}
The unitary operator on the left is simply the desired Hamiltonian evolution from \unionofstars along with a nonunitary factor $\exp(-\Gamma t/8)$ related to the definition of the effective Hamiltonian, 
\begin{equation} \mathcal{T} \left(e^{-i H_\mathrm{eff}t} \left\{ \prod_a\left[ \prod_{u \in B_a} \sigma_u^x\right]\right\}\right) = e^{-i \gamma(t) C'}e^{-\Gamma t/8}, \end{equation} 
where $\gamma(t) = t/t_{\gamma=1}$ with $t_{\gamma=1}$ the amount of time it takes to compile $\exp(-i C')$ (unitaries $\exp(-i \gamma C')$ are prepared by linearly scaling each step in the compilation by $t/t_{\gamma=1}$), while the jump operators define a trajectory-dependent effective initial state 
\begin{equation} \ket{\psi(0;T)} = \prod_i \mathcal{J}_{u_i}\ket{\psi(0)} \end{equation}
where we have set the physically-irrelevant global phase factor $\pm$ equal to unity.  To abbreviate notation we drop the time argument in $\gamma(t)$ below, until the final results are presented.

Following the approach of Ref.~\cite{foss2013nonequilibrium}, we will compute spin-spin correlation functions along a single trajectory, then average over the ensemble of trajectories to obtain exact results from the master equation (\ref{Lindbladian}).  We consider spin-spin correlation functions in terms of raising and lowering operators, $\sigma^+ = \ket{0}\bra{1}$ and $\sigma^- = \ket{1}\bra{0}$ respectively (which are components of $\sigma^y$ and $\sigma^x$), as well as $\sigma^z$. We will relate these to QAOA expectation values in the next subsection.  

Begin with the correlation function
\begin{align} \label{eq 20} \langle \sigma^+_u \sigma_v^+\rangle_T = \frac{\langle \psi(t;T) \vert \sigma^+_u \sigma_v^+ \ket{\psi(t;T)}}{\bra{\psi(t;T)}\psi(t;T)\rangle} = \bra{\psi_o(0;T)} e^{i\gamma C'}  \sigma^+_u \sigma_v^+ e^{-i \gamma C'} \ket{\psi_o(0;T)}
\end{align} 
where on the far right we have defined $\ket{\psi_o(0;T)} = \prod_i \sigma^z_{u_i}\ket{\psi(0)} $ as a normalized version of $\ket{\psi(0;T)}$; the nonnormalized factors $\sqrt{\Gamma/8}$ from the jump operators $\mathcal{J}$ will be included again later, in the probability definitions $\mathrm{Pr}(\mathcal{F})$ below, following the approach of Ref.~\cite{foss2013nonequilibrium}; see also the alternative derivation in Ref.~\cite{lotshaw2024exactly}. eqn.~(\ref{eq 20}) can be simplified by considering the middle term
\begin{equation} e^{i\gamma C'}  \sigma^+_u \sigma_v^+ e^{-i \gamma C'} = \ket{0_u,0_v}\bra{1_u,1_v} e^{i2\gamma\sum_{\mu \neq u,v} (c_{\mu u}'+c_{\mu v}')\sigma_\mu^z} \end{equation}
Then we have 
\begin{equation} \langle \sigma^+_u \sigma_v^+\rangle_T = \bra{\psi_o(0;T)} \left(\ket{0_u,0_v}\bra{1_u,1_v} e^{i2\gamma\sum_{\mu \neq u,v} (c_{\mu u}'+c_{\mu v}')\sigma_\mu^z} \right)\ket{\psi_o(0;T)} \end{equation}
Finally, noting that $\ket{\psi(0)} = \ket{+}^{\otimes n}$ we have 
\begin{equation} \langle \sigma^+_u \sigma_v^+\rangle_T = \frac{(-1)^{\mathcal{F}_u+\mathcal{F}_v}}{4} \prod_{\mu \neq u,v} \cos(2\gamma(c_{\mu u}' + c_{\mu v}')) \end{equation}
where $\mathcal{F}_u(T)$ and $\mathcal{F}_v(T)$ are the numbers of jump operators applied to $u$ and $v$ in the trajectory $T$. 

Similarly we have 

\begin{equation} \langle \sigma^+_u \sigma_v^-\rangle_T = \frac{(-1)^{\mathcal{F}_u(T)+\mathcal{F}_v(T)}}{4} \prod_{\mu \neq u,v} \cos(2\gamma(c_{\mu u}' - c_{\mu v}')) \end{equation}

\begin{equation} \langle \sigma^-_u \sigma_v^+\rangle_T = \frac{(-1)^{\mathcal{F}_u(T)+\mathcal{F}_v(T)}}{4} \prod_{\mu \neq u,v} \cos(2\gamma(-c_{\mu u}' + c_{\mu v}')) \end{equation}

\begin{equation} \langle \sigma^-_u \sigma_v^-\rangle_T = \frac{(-1)^{\mathcal{F}_u(T)+\mathcal{F}_v(T)}}{4} \prod_{\mu \neq u,v} \cos(2\gamma(-c_{\mu u}' - c_{\mu v}')) \end{equation}

Also
\begin{align} \langle \sigma^z_u \sigma^+_v\rangle_T & = \bra{\psi_o(0;T)} e^{i\gamma C'}  \sigma^z_u \sigma_v^+ e^{-i \gamma C'} \ket{\psi_o(0;T)} \nonumber\\ 
& = \frac{\bra{0_u} + \bra{1_u}}{\sqrt{2}} \sigma_u^ze^{i2\gamma c_{uv}\sigma_u^z} \frac{\ket{0_u} + \ket{1_u}}{\sqrt{2}}\frac{(-1)^{\mathcal{F}_v(T)}}{2} \prod_{\mu \neq u,v} \frac{e^{i2\gamma c_{uv}'} + e^{-i2\gamma c_{uv}'}}{2}\nonumber\\
& = i\sin(2\gamma c_{uv}')\frac{(-1)^{\mathcal{F}_v(T)}}{2}\prod_{\mu \neq u,v} \cos(2\gamma c_{\mu,v}'), \end{align} 
\begin{align} \langle \sigma^z_u \sigma^-_v\rangle_T = -i\sin(2\gamma c_{uv}')\frac{(-1)^{\mathcal{F}_v(T)}}{2}\prod_{\mu \neq u,v} \cos(2\gamma c_{\mu,v}') ,\end{align}
noting that there is phase factor missing in eqn.~(A8) of \cite{foss2013nonequilibrium} which is corrected above.  Finally 
\begin{equation} \langle \sigma_u^z \sigma_v^z \rangle_T = \bra{\psi(0)}\sigma_u^z \sigma_v^z \ket{\psi(0)}=0. \end{equation}

The final step is to average the correlation functions over all possible trajectories $T$.  Following \cite{foss2013nonequilibrium}, the dephasing terms for each qubit $u$ added randomly following a Poisson process with probability distribution $\mathrm{Pr}(\mathcal{F}_u) = e^{-\Gamma t/4} (\Gamma t/4)^\mathcal{F}_u/\mathcal{F}_u!$.  The exact spin-spin correlation functions come from averaging over this probability distribution, for example
\begin{equation} \langle \sigma^+_u \sigma^+_v\rangle = \sum_T \mathrm{Pr}(T) \langle \sigma^+_u \sigma^+_v\rangle_T = \sum_{\mathcal{F}_u,\mathcal{F}_v=0}\mathrm{Pr}(\mathcal{F}_u)\mathrm{Pr}(\mathcal{F}_v)\frac{(-1)^{\mathcal{F}_u+\mathcal{F}_v}}{4} \prod_{\mu \neq u,v} \cos(2\gamma(c_{\mu u}' + c_{\mu v}')) \end{equation} 
and using the identity
\begin{equation} \sum_{\mathcal{F}=0}^\infty \mathrm{Pr}(\mathcal{F})(-1)^\mathcal{F} = e^{-\Gamma t/2} \end{equation} 
we arrive at
\begin{align}
     \langle \sigma^+_u \sigma_v^+\rangle =\langle \sigma^-_u \sigma_v^-\rangle = \frac{e^{-\Gamma t}}{4} \prod_{\mu \neq u,v} \cos(2\gamma(c_{\mu u}' + c_{\mu v}'))\nonumber\\
\langle \sigma^+_u \sigma_v^-\rangle = \langle \sigma^-_u \sigma_v^+\rangle = \frac{e^{-\Gamma t}}{4} \prod_{\mu \neq u,v} \cos(2\gamma(c_{\mu u}' - c_{\mu v}')) \nonumber\\
\langle \sigma^z_u \sigma^\pm_v\rangle = \pm i\sin(2\gamma c_{uv}')\frac{e^{-\Gamma t/2}}{2}\prod_{\mu \neq u,v} \cos(2\gamma c_{\mu,v}')
\end{align}

\subsection{Translation into QAOA expectation values}

For QAOA we want to compute expectation values
\begin{equation} \bra{\psi(0)} e^{i \gamma C'} e^{i \beta B} \sigma_u^z \sigma_v^z e^{-i \beta B} e^{-i \gamma C'} \ket{\psi(0)} \end{equation} 
where $B = \sum_u \sigma_u^x$. The QAOA and Ising expectation values are related as 
\begin{equation} \langle \sigma_u^z \sigma_v^z \rangle_\mathrm{QAOA} =  \langle e^{i \beta B} \sigma_u^z \sigma_v^z e^{-i \beta B} \rangle \end{equation}
where the expectation on the right is with respect to the Ising evolution described earlier.  In terms of the Ising evolution we need to compute the expectation value of the operator
\begin{align} e^{i \beta B} \sigma_u^z \sigma_v^z e^{-i \beta B} & = e^{i \beta \sigma_u^x} \sigma_u^z e^{-i \beta \sigma_u^x} e^{i \beta \sigma_v^x} \sigma_v^z e^{-i \beta \sigma_v^x} \nonumber\\
& = (\cos(2\beta) \sigma_u^z + \sin(2\beta) \sigma_u^y)(\cos(2\beta) \sigma_v^z + \sin(2\beta) \sigma_v^y) \nonumber\\
& = \cos^2(2\beta) \sigma_u^z \sigma_v^z + \cos(2\beta)\sin(2\beta) (\sigma_u^y\sigma_v^z + \sigma_u^z\sigma_v^y) + \sin^2(2\beta) \sigma_u^y\sigma_v^y\end{align} 
Noting $\langle \sigma_u^z\sigma_v^z\rangle=0$, taking $\cos(2\beta)\sin(2\beta) = \sin(4\beta)/2$, and replacing $\sigma_u^y = -i\sigma^+_u + i \sigma^-_u$ we have
\begin{align} \langle e^{i \beta B} \sigma_u^z \sigma_v^z e^{-i \beta B} \rangle = & -i\frac{\sin(4\beta)}{2}(\langle \sigma_u^+\sigma_v^z\rangle - \langle \sigma^-_u\sigma_v^z\rangle + \langle\sigma_u^z\sigma_v^+\rangle -\langle \sigma_v^z\sigma_u^-\rangle ) \nonumber\\ 
& + \sin^2(2\beta) (\langle \sigma^+_u\sigma_v^-\rangle + \langle \sigma^-_u\sigma_v^+\rangle - \langle \sigma^+_u\sigma_v^+\rangle - \langle \sigma^-_u\sigma_v^-\rangle )\end{align}
From above 
\begin{align} -i (\langle \sigma_u^+\sigma_v^z\rangle - \langle \sigma^-_u\sigma_v^z\rangle) = -i\left[i \sin(2\gamma c_{uv}') \frac{e^{-\Gamma t/2}}{2}\prod_{\mu\neq u,v}\cos(2\gamma c_{\mu v}') + i \frac{e^{-\Gamma t/2}}{2}\sin(2\gamma c_{uv}') \prod_{\mu\neq u,v}\cos(2\gamma c_{\mu v}')\right] \nonumber\\ 
= e^{-\Gamma t/2}\sin(2\gamma c_{uv}') \prod_{\mu\neq u,v}\cos(2\gamma c_{\mu v}') \end{align}
\begin{align}
 -i (\langle \sigma_u^z\sigma_v^+\rangle - \langle \sigma^z_u\sigma_v^-\rangle) =  e^{-\Gamma t/2}\sin(2\gamma c_{uv}') \prod_{\mu\neq u,v}\cos(2\gamma c_{\mu u}')\end{align}
 \begin{align}\langle \sigma^+_u \sigma_v^-\rangle + \langle \sigma^-_u \sigma_v^+\rangle = \frac{e^{-\Gamma t}}{2} \prod_{\mu \neq u,v} \cos(2\gamma(c_{\mu u}' - c_{\mu v}'))\end{align}
 \begin{align}\langle \sigma^+_u \sigma_v^+\rangle + \langle \sigma^-_u \sigma_v^-\rangle = \frac{e^{-\Gamma t}}{2} \prod_{\mu \neq u,v} \cos(2\gamma(c_{\mu u}' + c_{\mu v}'))
 \end{align}
Then we have 
\begin{align} \langle e^{i \beta B} \sigma_u^z \sigma_v^z e^{-i \beta B} \rangle = \frac{\sin(4\beta)\sin(2\gamma c_{uv}')e^{-\Gamma t/2}}{2} \left( \prod_{\mu\neq u,v}\cos(2\gamma c_{\mu v}') + \prod_{\mu\neq u,v}\cos(2\gamma c_{\mu u}') \right) \nonumber\\
- \frac{\sin^2(2\beta)e^{-\Gamma t}}{2} \left(\prod_{\mu \neq u,v} \cos(2\gamma(c_{\mu u}' + c_{\mu v}')) - \prod_{\mu \neq u,v} \cos(2\gamma(c_{\mu u}' - c_{\mu v}'))\right) \end{align}

For a QAOA cost Hamiltonian $C = \sum_{u<v} c_{uv} \sigma^z_u \sigma^z_v$ (which may differ from the $C'$ depending on the compilation approach) the total cost expectation is
\begin{align} \label{C QAOA} \langle C \rangle_\mathrm{QAOA}  & = \sum_{u<v} \frac{c_{uv}\sin(4\beta)\sin(2\gamma(t) c_{uv}')e^{-\Gamma t/2}}{2} \left( \prod_{\mu\neq u,v}\cos(2\gamma(t) c_{\mu v}') + \prod_{\mu\neq u,v}\cos(2\gamma(t) c_{\mu u}') \right) \nonumber \\
 & - \sum_{u<v} \frac{c_{uv}\sin^2(2\beta)e^{-\Gamma t}}{2} \left(\prod_{\mu \neq u,v} \cos(2\gamma(t)(c_{\mu u}' + c_{\mu v}')) - \prod_{\mu \neq u,v} \cos(2\gamma(t)(c_{\mu u}' - c_{\mu v}'))\right) \end{align}
where we have explicitly noted the time argument in $\gamma(t) = t/t_{\gamma=1}$.  Terms related to  $\sigma^z \sigma^y$ decay at single-particle rates $\Gamma t/2$, while those related to $\sigma^y \sigma^y$ decay at twice that rate. In the limit $\Gamma \to 0$ and when $C = C'$, the expression (\ref{C QAOA}) agrees with the generic QAOA expectation value in eqn.~(14) of Ref.~\cite{ozaeta2022expectation}.

\end{document}